\documentclass[12pt,preprint]{aastex}
\def\apj{{ApJ}}

\def\apjs{{ApJS}}

\slugcomment{Draft, v. re-submitted to ApJ}
\shorttitle{{\it Suzaku} View of Local ULIRGs}
\shortauthors{Teng et al.}
\begin{document}

\title{{\it Suzaku} Observations of Local Ultraluminous Infrared Galaxies}

\author{Stacy H. Teng \altaffilmark{1,2}, Sylvain Veilleux \altaffilmark{2},
  Naohisa Anabuki \altaffilmark{3}, Charles D. Dermer \altaffilmark{4},
  Luigi~C.~Gallo \altaffilmark{5}, Takao~Nakagawa \altaffilmark{6},
  Christopher S. Reynolds \altaffilmark{2},  D.B. Sanders
  \altaffilmark{7}, Yuichi~Terashima \altaffilmark{8}, and Andrew~S.~Wilson
  \altaffilmark{2}}

\altaffiltext{1}{Contacting author: stacyt@astro.umd.edu.}
\altaffiltext{2}{Department of Astronomy, University of Maryland,
  College Park, MD 20742, U.S.A.}
%%\altaffiltext{2}{Contacting author.}
\altaffiltext{3}{Department of Earth and Space Science, Osaka
  University, 1-1 Machikaneyama Toyonaka, Osaka 560-0043, Japan}
\altaffiltext{4}{Space Science Division, Code 7653, U.S. Naval
  Research Laboratory, Washington, D.C. 20375-5352, U.S.A.}
\altaffiltext{5}{Department of Astronomy and Physics, Saint Mary's University,
Halifax, NS B3H 3C3, Canada}
\altaffiltext{6}{Institute of Space and Astronautical Science, Japan Aerospace
Exploration Agency, 3-1-1 Yoshinodai, Sagamihara, Kanagawa 229-8510, Japan}
\altaffiltext{7}{Institute for Astronomy, University of Hawaii, 2680
  Woodlawn Drive, Honolulu, HI 96822, U.S.A.}
\altaffiltext{8}{Department of Physics, Ehime University, Matsuyama,
  Ehime 790-8577, Japan}

\begin{abstract}

  We report the results from our analysis of {\it Suzaku} XIS
  (0.5--10~keV) and HXD/PIN (15--40~keV) observations of five
  well-known local ULIRGs: {\em IRAS} $F$05189--2524, {\em IRAS}
  $F$08572+3915, Mrk~273, PKS~1345+12, and Arp~220.  The XIS
  observations of $F$05189--2524 and Mrk~273 reveal strong iron lines
  consistent with Fe~K$\alpha$ and changes in spectral shapes with
  respect to previous {\it Chandra} and {\it XMM-Newton} observations.
  Mrk~273 is also detected by the HXD/PIN at $\sim$1.8-$\sigma$.  For
  $F$05189--2524, modeling of the data from the different epochs
  suggests that the change in spectral shape is likely due to the
  central source switching off, leaving behind a residual reflection
  spectrum, or an increase in the absorbing column.  An increase in the covering fraction of the absorber can
  describe the spectral variations seen in Mrk~273, although a
  reduction in the intrinsic AGN luminosity cannot be formally ruled
  out.  The {\it Suzaku} spectra of Mrk~273 are well fit by a $\sim$94\% covering
  fraction model with a column density of $\sim$10$^{24}$~cm$^{-2}$.
  The absorption-corrected log[$L_{\rm 2-10~keV}$/$L_{\rm IR}$] ratio
  is consistent with those found in PG Quasars.  The change
  in the spectral shape on a time scale of a few years implies that
  the absorbing matter must be near the AGN ($\sim$1~pc).  The
  0.5--10~keV spectrum of PKS~1345+12 and Arp~220 seem unchanged from
  previous observations and their hard X-ray emission is not
  convincingly detected by the HXD/PIN.  The large column density
  derived from CO observations and the large equivalent width of an
  ionized Fe line in Arp 220 can be reconciled by an ionized
  reflection model.  $F$08572+3915 is undetected in both the XIS and
  HXD/PIN, but the analysis of unpublished {\em Chandra} data provides
  a new measurement at low energies.

\end{abstract}


\section{Introduction}
\label{sec:intro}

Ultraluminous Infrared Galaxies (ULIRGs) are galaxies with $L_{\rm
  IR}$ = $L_{8-1000 \mu {\rm m}}$ $\ge$ 10$^{12} L_{\odot}$,
  equivalent to the minimum bolometric luminosity of QSOs.  The
  discovery of a large population of local ULIRGs is one of the most
  important legacies of the {\em IRAS} satellite.  At luminosities
  above 10$^{12} L_{\odot}$, the space density of ULIRGs in the local
  universe is greater than that of optically selected quasars with
  similar bolometric luminosity by a factor of $\sim$1.5 (Sanders \&
  Mirabel 1996). Thus ULIRGs represent the most common type of galaxy
  at these high luminosities. Systematic ground-based and space-based
  optical and near-infrared imaging studies have shown that local
  ULIRGs are almost always undergoing mergers (e.g., Sanders et al.
  1988; Veilleux et al. 2002, 2006).  Sanders et al. (1988) suggested
  that these objects represent a dust-enshrouded phase that eventually
  evolves into optically-selected quasars. If this is true, ULIRGs
  take on a fundamental importance for the origin and evolution of
  quasars.

The primary energy source --- active galactic nucleus (AGN) versus
starburst activity --- of ULIRGs is still under debate.  Optical and
infrared observations show that most local ULIRGs are dominated by
starbursts, but about 30\% show evidence of AGNs \citep{vei99a,
  vei99b}.  Furthermore, the ``warm'' infrared colors ($f_{25~\mu {\rm
    m}}$/$f_{60~\mu {\rm m}}$ $>$ 0.2) and quasar-like spectra of the
more luminous objects imply that active nuclei are significant in
these objects (e.g., Veilleux et al. 1995, 1999a,b, 2008; Genzel et
al. 1998; Tran et al. 2001; Armus et al. 2007).  These results would
be definitive if it were not for the possibility of severe
obscuration. 

Galaxy mergers cause massive in-falls of gaseous material towards the
center of ULIRGs (e.g., Rupke, Veilleux, \& Baker 2008), and column
densities $\ga$ 10$^{24}$ cm$^{-2}$ have been deduced from CO
measurements (e.g., Downes \& Solomon 1998; Evans et al. 2002).
Observations at UV, optical, near-infrared, and even mid-infrared wavelengths may
therefore not always probe the cores of these objects. High resolution
($\ll$ 1''), high frequency radio observations can penetrate high
columns and are an excellent probe of whether an AGN is present (e.g.,
Lonsdale, Smith, \& Lonsdale 1993).  However, the bolometric
luminosity in the radio band is insignificant, so radio data cannot
test whether accretion onto a supermassive black hole (SMBH) is the dominant energetic process.
We are left with hard X-rays.

Recent surveys with {\it XMM-Newton} and {\it Chandra} (e.g. Ptak et
al. 2003; Franceschini et al. 2003; Teng et al. 2005) have found that
$\sim$40\% of observed ULIRGs show signatures of AGNs.  The
observed 2--10~keV luminosity of the surveyed sample is
$\sim$10$^{40}$--10$^{43}$~ergs~s$^{-1}$, with a majority of the
sources having luminosities below 10$^{42}$~ergs~s$^{-1}$.  While the
ratio log[$L_{\rm 2-10~keV}$/$L_{\rm IR}$] is small in nearby ULIRGs
(from $-$4 to $-1$; e.g., Teng et al. 2005), this is not much smaller
than that found in radio-quiet QSOs (from $-$3 to $-$1).  Moreover,
absorption may be a factor even at these energies.  If the absorbing
column exceeds $\sim$10$^{24}$ cm$^{-2}$, the primary continuum
emission is suppressed significantly by absorption and Compton
down-scattering.  Thus observations at $\ga$ 10 keV are best to detect
Compton-thick AGNs. The sensitivity of {\it Suzaku} at these high
energies is well suited for the study of highly obscured sources like
ULIRGs.  

In this paper, we present {\it Suzaku} XIS (0.5--10~keV) and HXD/PIN
(15--40~keV) observations of five well-known local ULIRGs. In \S 2, we
discuss our sample. In \S 3, we report the observations and describe
the methods we used to reduce the data. In \S 4, the results from our
spectral analysis of the {\em Suzaku} data are discussed. In \S 5, we
combine the {\em Suzaku} data with earlier published and unpublished
{\em XMM-Newton} and {\em Chandra} data to fine tune our spectral
models.  The results of this study are summarized in \S 6.  Throughout
this paper, we adopt the cosmology of
H$_0$=75~km~s$^{-1}$~Mpc$^{-1}$, $\Omega _{\rm M}$=0.3, and $\Omega
_\Lambda$=0.7.

\section{Sample}
\label{sec:sample}

The five ULIRGs in the present study are $F$05189--2524,
$F$08572+3915, Mrk~273, PKS~1345+12, and Arp~220. They were selected because they
have readily available {\em Suzaku} data, either from our own program
(PI: Veilleux) or from the public archive.
They are among the nearest, brightest, and best-studied ULIRGs in the
{\em IRAS} Bright Galaxy Survey \citep{sanders03}.
Table~\ref{tab:sample} lists the basic properties of these sources. Here
we briefly review the relevant literature on each source.

$F$05189--2524 is an unresolved late stage merger surrounded by tidal
debris (Veilleux et al. 2002, 2006) with ``warm'' infrared colors.  It
is optically classified as a Seyfert~2, but near-infrared spectroscopy
of this source reveals the presence of an obscured broad line region
(BLR) at Pa$\alpha$ (Veilleux et al. 1999a, b).  Spectra from previous
{\it ASCA} \citep{risaliti00, severgnini}, {\it XMM-Newton} \citep{iman04},  and {\it Chandra} \citep{ptak}
observations flatten out above 2~keV and are best fit by an absorbed
power law with $N_{\rm H} \sim 0.5-1 \times $10$^{23}$ cm$^{-2}$, $\Gamma
\sim$1.0--1.9, and a thermal component with kT$\sim$0.1--0.9~keV.  The
absorption-corrected 2--10~keV luminosity derived from the published {\it
  XMM-Newton} and {\it Chandra} data is $\sim$10$^{43}$ ergs s$^{-1}$.

$F$08572+3915 is another ``warm'' ULIRG, consisting of a pair of
interacting galaxies with nuclear separation of $\sim$6~kpc \citep[and
references therein]{vei02}. The northwestern nucleus is classified as
a LINER.  A 2-cm radio core coincides with the northwestern nucleus
\citep{nagar}.  However, only an upper limit on its 2--10~keV flux
exists in the literature -- $7.6 \times 10^{-12}$ ergs cm$^{-2}$
s$^{-1}$ -- from the {\it HEAO} satellite \citep{heao}.

Mrk~273 is in the early phase of a merger where the nuclei are
separated by 680 pc \citep{vei02} and is optically classified as a
Seyfert~2 \citep{vei99a}.  A [Si~VI] 1.96 $\mu$m feature, a strong
indicator of AGN activity, is detected in this ``cool'' ULIRG
\citep{vei99b}.  Its radio flux falls above the radio-to-FIR
correlation of starbursts and a bright AGN-like radio core is detected
on VLBA scale in this object \citep{lonsdale}.  The {\it Chandra}
0.5--2~keV X-ray spectrum of Mrk~273 is best explained with a MEKAL
plasma with kT $\sim$1.3~keV and its 2--10~keV spectrum is best fit by
an absorbed power law with $\Gamma \sim$1.0 and $N_{\rm H}
\sim$10$^{23}$ cm$^{-2}$ \citep{ptak}.  The flat slope of the
2--10~keV continuum may be a result of reflection, but the {\it
  Chandra} observation of \citet{ptak} shows a very weak Fe~K$\alpha$
emission line with an equivalent width of $\sim$0.09--0.44~keV.  Comparisons of observations by {\it ASCA}, {\it BeppoSAX}, and {\it Chandra} show that this source exhibits possible long term flux variability \citep{xia}. 

PKS~1345+12 is yet another ``warm'' ULIRG.  It is in the early stage
of a merger with two nuclei separated by 4.0~kpc \citep{vei02}.
Optically classified as a Seyfert~2 galaxy \citep{bgs, kim98},
infrared observations by \citet{vei97} suggest a buried BLR at
Pa$\alpha$.  Observations by \citet{evans99} have shown that the
eastern nucleus has colors consistent with reddened starlight while
the the western nucleus has extremely red colors indicative of an
optical quasar.  The western nucleus is also coincident with peak CO
emission \citep{evans99}, a radio core \citep{nagar}, and 0.5--8~keV
X-ray emission \citep{iman04}.  According to \citet{iman04}, the {\it
Chandra} continuum is consistent with that of an absorbed AGN (power
law $\Gamma$=1.8 with $N_{\rm H} \sim$4.5$\times 10^{22}$ cm$^{-2}$).
The {\it Chandra} data also showed a narrow Fe~K$\alpha$ emission line
with an equivalent width of $\sim$0.13~keV.

Finally, Arp~220 is by far the best-studied ULIRG due to its vicinity.
Optically classified as a LINER, this ``cool'' ULIRG is also an early
merger with nuclei separated by $\sim$0.4~kpc \citep{vei02}.
Multi-wavelength data suggest the presence of a black hole in the
western nucleus \citep[and references therein]{downes07}.  Previous
{\it Chandra} observations detected both nuclei as well as extended
soft X-ray emission from lobes and plumes that extend beyond the
optical galaxy \citep{arp220a, arp220b}.  The full {\it Chandra} band
nuclear spectrum is best
fit by a thermal MEKAL component with $kT$ $\sim$0.8~keV and a flat
power law with $\Gamma \sim$1.1 absorbed by a column of
$\sim$10$^{21}$ cm$^{-2}$ \citep{ptak}.  Analysis of {\em XMM-Newton}
data by \citet{iwasawa} suggests the presence of an Fe~K emission line
with equivalent width of $\sim$1.9~keV emanating from the western
nucleus.

\section{Observations and Data Reduction} 
\label{sec:obs}

The details of the {\it Suzaku} observations are listed in
Table~\ref{tab:obs}.  Three of the sources ($F$05189--2524, Mrk~273, and PKS~1345+12) were of our own program (PI:
Veilleux, Anabuki was co-PI for $F$05189--2524) and the other two,
$F08572+3915$ (PI: Gallagher) and Arp~220 (PI: the {\it Suzaku}
Science Working Group, or SWG), were
downloaded from the public archive.  All of the observations were performed at the HXD aim point to
increase the sensitivity of the HXD.   The analysis of the data on $F$05189--2524, $F$08572+3915, Mrk~273, and Arp~220 was performed with version 6.3.1 of HEASoft and CALDB version 20071016.  PKS~1345+12 was observed at the end of Cycle~2 and after the completion of the data analysis on the other objects in the sample.  Thus, the analysis of the data on PKS~1345+12 was performed with the more up-to-date version 6.4 of HEASoft and CALDB version 20080401.  

\subsection{XIS Data Reduction}
\label{sec:xisobs}

The XIS data reduction followed the guidelines provided in the {\it
  Suzaku} Data Reduction Guide\footnote{See
  http://heasarc.gsfc.nasa.gov/docs/suzaku/analysis/abc/.}.  The data
were screened following the version 2 data screening criteria.
Due to nearby field sources, cleaned events for all objects except PKS~1345+12 were then extracted in circular regions with 1$\arcmin$
radii (the minimum recommended region size) centered on the targets
wherever possible.  The PKS~1345+12 field is rather empty, so the extraction region has a radius of $\sim$3.1$\arcmin$.  Since the extraction regions are so large, {\it
  Chandra} ACIS data were used as a check to ensure no other X-ray
sources were included in the extraction regions. 
Background events were extracted in same-sized nearby source-free regions.  The source spectra were binned to at
least 50 counts bin$^{-1}$ for PKS~1345+12 and at least 15 counts
bin$^{-1}$ for the others so that $\chi ^2$ statistics can be used
when modeling the spectra.

The response (RMF) and auxiliary (ARF) files were produced using {\tt
  xisrmfgen} and {\tt xissimarfgen}.  The ARF files were generated
assuming 400,000 incident photons and the default grid spacing.  

In the modeling of the XIS data, the XIS1, XIS2, and XIS3 detectors
are assumed to have the data processing version 2.0
cross-normalization factors of 1.065, 1.035, and 1.067 with respect to
the XIS0 detector, respectively\footnote{See:
  ftp://legacy.gsfc.nasa.gov/suzaku/doc/xrt/suzakumemo-2007-11.pdf.}.

\subsection{HXD/PIN Data Reduction}
\label{sec:hxdobs}

The HXD/PIN data were reduced following the guidelines provided by the
{\it Suzaku} team.  The HXD/PIN spectra for both sources were
extracted after the selection of good time intervals (the ``ANDed
GTI'' from both the data and the non-X-ray background provided by the
{\it Suzaku} HXD team).  Dead-time corrections (on the order of about
5\%) were applied to the extracted source spectra. The extracted
spectra were binned using {\tt grppha}\footnote{We binned the data
  using group 0 31 2 32 63 4 64 95 8 96 127 32 128 255 64.  Each set of three numbers represent the channel range grouped and the number of channel bins in each group.
This choice of grouping was used so that each bin contains approximately the same number of photons.}.

The response files were provided by the HXD team.  Due to changes in
the bias voltages and the threshold over time, we used the first-epoch
response file for observations of $F$05189--2524, $F$08572+3915, and
Arp~220, second-epoch for Mrk~273, and fourth-epoch for PKS~1345+12.

As its name implies, the non-X-ray backgrounds (NXB) from charged
particles modeled by the
HXD team do not include the cosmic X-ray background (CXB) which peaks
within the energy range of the HXD/PIN.  The CXB for each galaxy was
modeled following a recipe\footnote{See: http://heasarc.gsfc.nasa.gov/docs/suzaku/analysis/pin\_cxb.html.} provided by the HXD team.  The simulated
CXB is approximately 5\% of the NXB.  The NXB and CXB were added
together using {\tt mathpha} to provide total backgrounds for the
HXD/PIN data.

At the time of this writing, the accuracy of the processing version
2.0 HXD/PIN background model is 3.8\%\footnote{This is the 1-$\sigma$
  statistical plus systematic error in the 15--40~keV band for a net
  integration time of 10~ksec.  See:
  ftp://legacy.gsfc.nasa.gov/suzaku/doc/hxd/suzakumemo-2007-09.pdf.}.
However, a $\sim$10\% offset in the version 2.0 background model was
discovered for data taken between March and May of 2006 due to changes
in the PIN observing mode.  For the affected data ($F$05189--2524 and
$F$08572+3915), processing version 1.2 HXD/PIN background were used as
recommended by the {\it Suzaku} team.  Dead time corrections were also
performed on the backgrounds for the affected data\footnote{See:
  http://www.astro.isas.ac.jp/suzaku/analysis/hxd/v1/pinnxb/}.  The
reproducibility of the version 1.2 background is between 5 and
10\%\footnote{See:
  ftp://legacy.gsfc.nasa.gov/suzaku/doc/hxd/suzakumemo-2006-43.pdf.}.
For the modeling of the data, the cross-normalization of the PIN with
respect to XIS0 is 1.16 and 1.13 for the version 1.2 and 2.0
backgrounds, respectively\footnote{See:
  ftp://legacy.gsfc.nasa.gov/suzaku/doc/xrt/suzakumemo-2006-40.pdf for
  version 1.2 backgrounds and suzakumemo-2007-11.pdf for version 2.0
  backgrounds.}.

\section{{\it Suzaku} Results}
\label{sec:results}

The spectra were analyzed using the XSPEC package (version 11.3.2ag).
The energy range of the detectors were limited to 0.5--10~keV for the
XIS detectors and 15--40~keV for the PIN detector to avoid calibration
problems.  All the errors in the parameters are at the 90\% level for
one parameter of interest ($\Delta \chi ^2 = 2.7$).  

In the modeling of the spectra, a simple power-law distribution
absorbed only by the Galactic column is first assumed.  If the model
is not a satisfactory fit to the data, then more components are added
on to the model until a satisfactory fit is achieved.  The continuum
model components considered in this paper are: (1) an absorbed
power-law model which represents emission from an AGN (model
parameters fitted include a column density, $N_{\rm H}$, a power law
index, $\Gamma$, and a normalization factor), (2) a thermal MEKAL
model representing emission from stars or galactic winds (with a gas
temperature, $kT$, and a normalization factor), (3) a partial-covering
fraction model where some fraction, $f_{\rm cover}$, of the intrinsic
nuclear radiation is absorbed and the rest goes through unimpeded.
The other parameters fitted are the equivalent column density, $N_{\rm
H}$, $\Gamma$, and a normalization factor.  (4) a scattering model
where an absorbed power law represents the direct transmitted
component and an unabsorbed power law of the same photon index
represents a scattered component ($N_{\rm H}$, $\Gamma$, and a
normalization factor for each of the direct and scattered components),
(5) a neutral reflection model\footnote{The reflection component is represented by the {\tt PEXRAV} model in {\tt XSPEC}.  For the spectral fitting, the metal abundance, iron abundance, and inclination angle are fixed at the default value.  The relative reflection parameter is fixed at --1, thus modeling only the reflection component.  No energy cut off is assumed.  Therefore, only $\Gamma$ and the normalization factor are free parameters.} with an unabsorbed power law
representing the direct component and a reflection component with the
same photon index as the direct component ($\Gamma$ and a
normalization factor for each of the reflected and transmitted
components), and finally, (6) a pure ionized reflection model as
proposed by \citet{reflion} without any direct transmitted component
($\Gamma$ and a normalization factor).

\subsection{{\em IRAS}~$F$05189--2524}
\label{sec:f05189}

\subsubsection{XIS Spectrum}
\label{sec:f05189xis}

\begin{figure}
\figurenum{1}
\epsscale{1}
\plotone{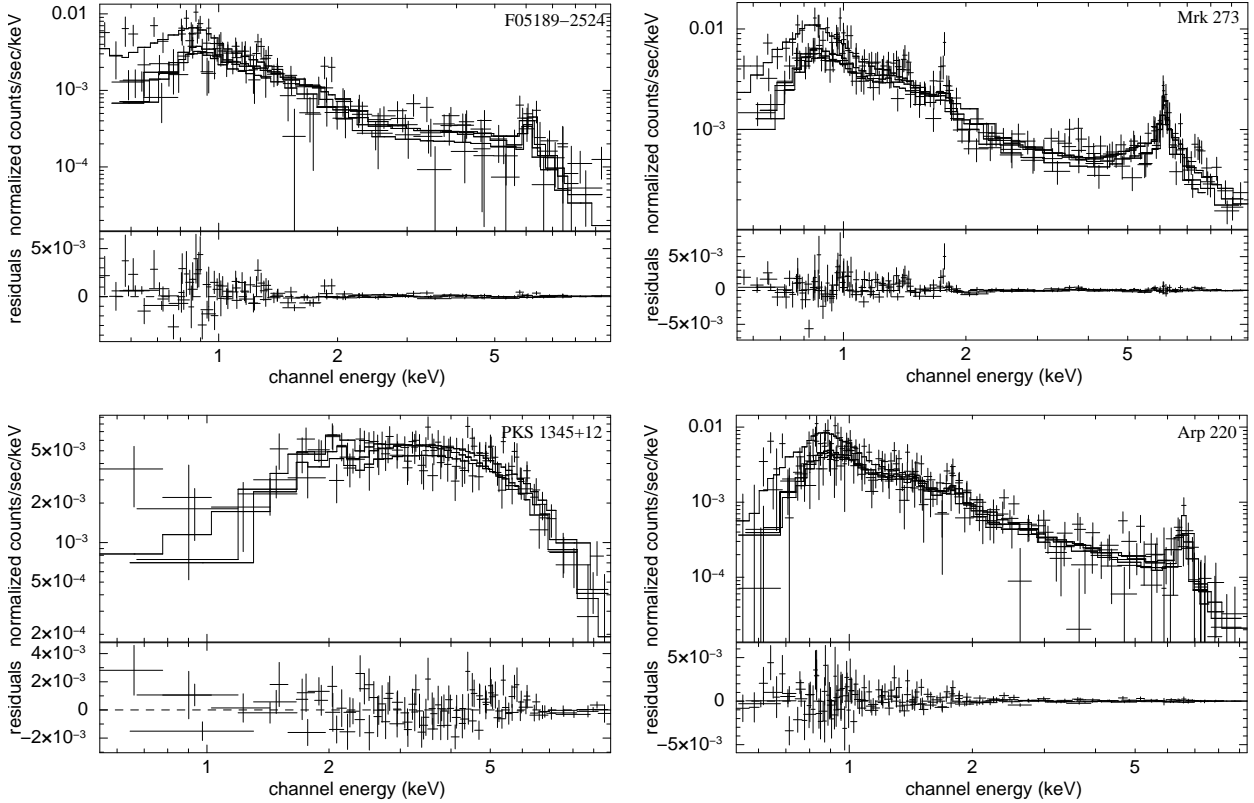}
\caption{Spectra of $F$05189--2524 (top left), Mrk~273 (top right), PKS~1345+12 (bottom left) and
  Arp~220 (bottom right) with their respective best-fit models for all four
  XIS detectors.  $F08572+3915$ is not detected by the XIS.  The data are binned to at least 50 counts bin$^{-1}$ for PKS~1345+12 and at least 15 counts bin$^{-1}$ for the others.
  The horizontal axis is energy in the observer's frame. In
  $F$05189--2524, Mrk~273, and Arp~220, an emission line is detected
  near 6.4--6.7~keV, consistent with emission arising from neutral or
  ionized iron.  The same emission line also appears to be present in
  PKS~1345+12, but the detection is not statistically significant
  ($\Delta \chi^2$=2.0 for $\Delta$d.o.f=2).
  Other than PKS~1345+12, the 0.5--2~keV spectrum for each object has a thermal component with temperatures of
  $\sim$0.7--0.8~keV, consistent with earlier results on ULIRGs. 
  The best-fit parameters are listed in  Table~\ref{tab:fitxis}.  }
\label{fig:xisfig}
\end{figure}

The XIS spectrum of $F$05189--2524 is shown in
Figure~\ref{fig:xisfig}.  First, we find that it is well-modeled by a
simple power-law model plus a MEKAL component with abundances fixed at
solar all modified by Galactic absorption ($\chi^2_\nu = 1.57$ for 124
degrees of freedom).  An additional Gaussian component is needed
($\Delta \chi ^2$=22.8 for a change in d.o.f. of 3) to
reproduce the emission line at $\sim$6.4~keV ({\em i.e.} the
Fe~K$\alpha$ line; equivalent width, or EW, $\sim$1.84~keV).  While
the spectrum shows excess emission at around 1.3 and 1.9~keV (Si~XIII
emission features), the addition of a second and a third Gaussian
component to the model does not significantly improve the fit ($\Delta
\chi ^2$=0.05 for a change in d.o.f. of 3).  The excess
emission at 1.9~keV may be a calibration feature associated with
absorption edges from the mirror.  The relatively flat
spectral index and the large equivalent width of the iron line suggest
that the spectrum may be reflection dominated.  The addition of a
reflection component to the continuum model results in a better fit
($\chi^2_\nu = 1.49$ for 123 d.o.f.).  The $F$-test
probability for the addition of one parameter (the normalization of
the reflected component) is relatively small, at $1.6 \times 10^{-2}$,
implying that it is reasonable to add the extra component.  While the
spectral index required for this fit, $\Gamma = 2.68^{+0.30}_{-0.13}$,
is unusually steep for an AGN where the canonical value for $\Gamma$
is $\sim$1.8, it is not too different from the range seen in PG
quasars \citep[1.3--2.48;][]{p05}.  The best-fit reflection model
parameters to the XIS spectrum are listed in Table~\ref{tab:fitxis}
and the best-fit model is shown with the spectrum in
Figure~\ref{fig:xisfig}.

\subsubsection{HXD/PIN Data}
\label{sec:f05189hxd}

$F$05189--2524 is undetected by the HXD/PIN.  As mentioned in
\S~\ref{sec:hxdobs}, the documentation provided by the {\it Suzaku}
team state that the reproducibility of the version 1.2 HXD/PIN
backgrounds are on average $\sim$5--10\%.  Here we estimate the
systematic and statistical error of the HXD/PIN data for our specific
observation.  The standard deviation of the histogram of the
15--40~keV residuals from the 1-day (40 ksec) observations in the
documentation imply that the systematic error of the NXB is
$\sim$2.4\%.  Considering the net HXD/PIN integration time and the
15--40~keV total background count rate for observations with
integration times longer than 40 ksec, the statistical error of the
$F$05189-2524 observation is $\sim$0.9\%.  Thus, the total 1-$\sigma$
error (systematic and statistical errors added in quadrature) for the
HXD/PIN observation of $F$05189-2524 is $\sim$2.6\%.  Therefore, the
15--40~keV 3-$\sigma$ detection limit for this observation is
$\sim$0.020 counts per second (cps).

\subsection{{\em IRAS}~$F$08572+3915}
\label{sec:f08572}

\subsubsection{XIS Non-Detection and {\em XMM-Newton} and {\em Chandra} Data}
\label{sec:f08572xis}

As mentioned previously in \S~\ref{sec:sample}, there has never been a
published X-ray detection of $F$08572+3915 in the X-ray energy range.  It is
also undetected by the XIS. 
A constraint on the limiting flux of $F$08572+3915 can be derived from the background.
The average background count rate from the XIS front illuminated detectors
(XIS0, 2, 3) is $\sim$1.3$\times 10^{-3}$ cps.
Assuming an ideal, flat background ($\Gamma \sim$1), the {\tt
WebPIMMS}\footnote{http://heasarc.gsfc.nasa.gov/Tools/w3pimms.html} application was used to estimate the observed flux of the background
spectrum.  The application yields a flux of $\sim$4$\times 10^{-14}$
ergs s$^{-1}$ cm$^{-2}$, more than two orders of
magnitude better than the upper limit placed on this object by {\it
  HEAO}.

$F$08572+3915 was observed with both {\em XMM-Newton} and {\it
  Chandra}, but the results of the analysis were never published. 
$F$08572+3915 is undetected in the archived {\em XMM-Newton}
data (ObsID: 0200630101, PI: Imanishi, t$_{\rm effective}$=13 ksec).
But the lower background and better spatial resolution of the {\it
Chandra} ACIS-S archival data (ObsID: 6862, PI: Komossa, t$_{\rm
effective}$=15 ksec) provide a conclusive detection of nine counts in
the 0.5--8~keV band.  The {\it Chandra} detection coincides with the
optical northwestern nucleus ({\it i.e.} the LINER/Seyfert~2
nucleus).  Assuming a canonical AGN power law spectrum ($\Gamma$=1.8)
modified by Galactic absorption for the Seyfert~2 nucleus, the detected count rate of $F08572+3915$ in the {\it
  Chandra} band corresponds to an observed 0.5--10~keV flux of $\sim 5 \times 10^{-15}$ ergs
s$^{-1}$ cm$^{-2}$.  This estimate is consistent with the upper limit derived from the XIS observations.   

The {\it Chandra} detection implies that the Seyfert nucleus may be a weak X-ray source or a heavily obscured AGN.
If we
assume no intrinsic absorption, the 0.5--10~keV flux from the {\it
  Chandra} observation of this object implies an intrinsic luminosity
of $\sim$5$\times 10^{40}$ ergs s$^{-1}$.  This luminosity falls
within the range of X-ray luminosities of LINERs as measured by
\citet{llagn} and is consistent with LINERs being powered by
low-luminosity AGNs.

However, like many other LINER~2s \citep{llagn}, $F$08572+3915 is
likely to be affected by absorption. Infrared observations of this
object do show strong absorption features and weak PAH emission
characteristic of deeply buried AGNs with a line-of-sight extinction
of $A_V \geq$~78 mag.\ (Imanishi et al. 2006; Armus et al. 2007;
Veilleux et al. 2008).  A CO-based estimate of the column density by
\citet{evans02} is in the range of $\sim$3--10$\times 10^{24}$
cm$^{-2}$.  The hardness ratio of the $F$08572+3915 detection is 0.56.
Following the hardness ratio method presented in \citet{teng}, a
single power law with an estimated photon index of $\sim$--0.43 would
fit the data.  The inverted power law spectrum may be an indication of
a large column density since the softer 0.5--2~keV photons are more
readily affected by absorption.  The estimated 0.5--10~keV flux from
the hardness ratio method, which assumes Galactic absorption only, is
$2.61 \times 10^{-13}$ ergs s$^{-1}$ cm$^{-2}$, corresponding to a
luminosity of $\sim 2 \times 10^{42}$ ergs s$^{-1}$.

\subsubsection{HXD/PIN Data}
\label{sec:f08572hxd}

$F$08572+3915 is not detected by the HXD/PIN above the total
background.  Following the procedure outlined in \S~\ref{sec:f05189hxd}, the HXD/PIN
3-$\sigma$ detection limit was used to approximate the limiting count
rate and observed flux in the 15--40~keV band.  The 3-$\sigma$ upper limit on the count
rate for $F$08572+3915 in the HXD/PIN band is $\sim$0.020 cps.

\subsection{Mrk~273}
\label{sec:m273}

\subsubsection{XIS Spectrum}
\label{sec:m273xis}

Mrk~273 is detected by the XIS.  We first tried to model the XIS
spectrum with a MEKAL component describing the starburst and a heavily
absorbed power-law distribution which does not fit the 2--10~keV
spectrum properly.  Rather, the XIS spectrum is well modeled by the
scattering model plus a MEKAL component with abundances fixed at solar
and all components modified by Galactic absorption.  Approximately 9\%
of the intrinsic AGN emission is scattered.  An additional Gaussian
component is needed to model the emission line at $\sim$6.4~keV
(EW$\sim$0.56~keV).  While the spectrum shows excess emission at
around 1.9~keV (Si~XIII emission), the addition of a second Gaussian
component to the model does not significantly improve the fit ($\Delta
\chi ^2$=1 with $\Delta$d.o.f.=3).  Again, the feature at 1.9~keV may
be a calibration feature associated with absorption edges from the
mirror.  A reflection model was also tested on the XIS spectrum, but
this results in a worse fit ($\chi _\nu ^2 =1.52$ for 187 d.o.f.) than
the scattering model ($\chi _\nu ^2 =1.25$ for 187 d.o.f.).  The
best-fit parameters to the XIS spectrum are listed in
Table~\ref{tab:fitxis} and the XIS spectrum with the best-fit model is
shown in Figure~\ref{fig:xisfig}.

\subsubsection{ HXD/PIN Data \& Contaminants}
\label{sec:m273amb}

\begin{figure}
\figurenum{2}
\epsscale{0.5}
\plotone{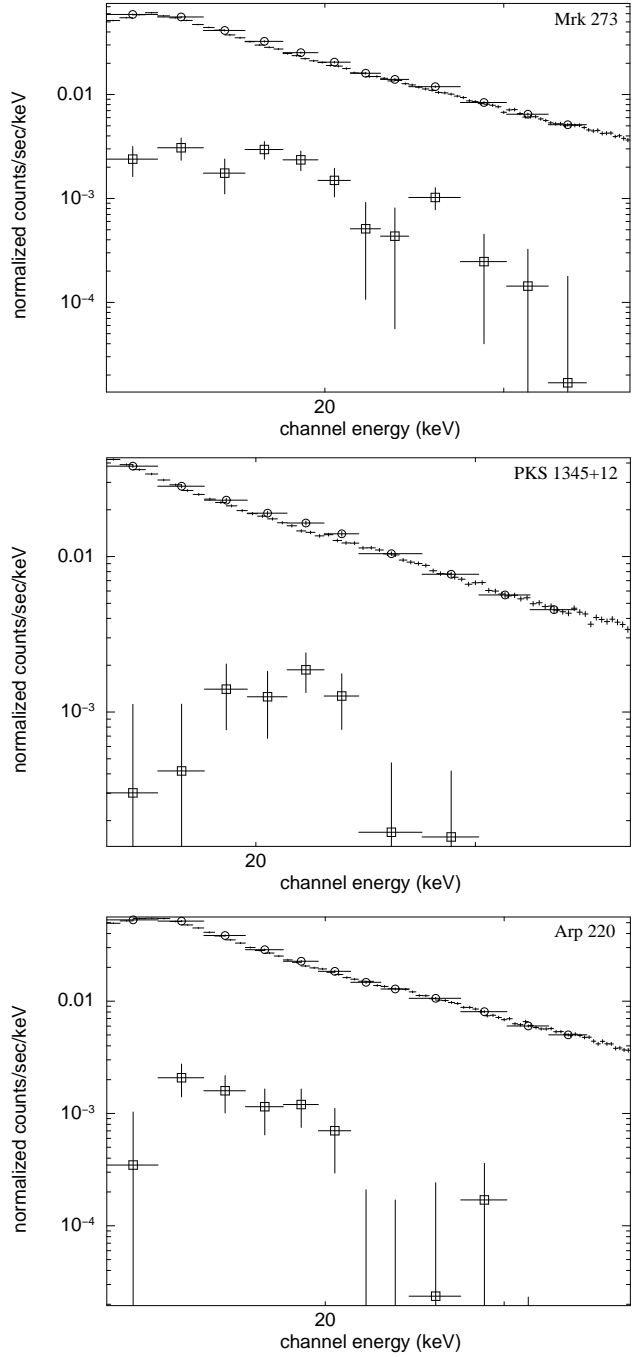}
\caption{HXD/PIN spectrum (open circles), together with the background
  spectrum (NXB+CXB; small crosses) and the net spectrum (open
  squares) of Mrk~273, PKS~1345+12, and Arp~220. The net spectrum for
  Mrk~273, PKS~1345+12, and Arp~220 is 5.4\%, 3.4\%, and 2.2\% above the background,
  respectively.  Mrk~273 is the only source that is
  detected above the background beyond the uncertainties of the background
modeling (at the 1.8-$\sigma$ level).  The spectra of $F$05189--2524 and
  $F$08572+3915 do not lie above the background, so they are not shown here.}
\label{fig:speccomp}
\end{figure}

Mrk~273 is marginally detected by the HXD/PIN.  The net (observed
minus background) HXD/PIN spectrum of Mrk~273 is $\sim$5.4\% of the
total (NXB+CXB) background.  Figure~\ref{fig:speccomp} is a comparison of
the spectrum, total background, and net spectrum.  The current
best estimate for the error in the version 2.0 NXB background is 3.8\%
for observations with net integration time of 10 ksec.  We estimate
the error on our Mrk~273 observation following the steps outlined in
\S\ref{sec:f05189hxd}.  The total systematic plus statistical error
for the Mrk~273 observation is $\sim$3.0\%, implying the HXD/PIN
detection of Mrk~273 is at $\sim$1.8-$\sigma$ (or with $\sim$93\%
confidence).

Because the PIN is not an imaging detector and has a very large field
of view (34$\arcmin \times 34\arcmin$, $\sim$3.5 times the area of the
XIS field of view), there may be sources of contamination in the
HXD/PIN signal.  While hard X-ray and soft gamma ray catalogs
(e.g. the INTEGRAL reference catalog; Ebisawa et al. 2003) do not list
any sources of comparable flux within the PIN field of view and energy
range, most of these flux values are based on modeling of previous
observations at lower energies.  There remains a possibility that the
signal may come from a previously unobserved nearby source.  Three
likely candidates are detected within the field of view of the XIS.
Mrk~273x is an unabsorbed Seyfert~2 galaxy at a redshift of 0.458
located approximately 1.2\arcmin\ northeast of Mrk~273.  The XIS
spectrum of Mrk~273x is well fit by a power-law component modified
only by Galactic absorption ($\Gamma \sim$1.69$^{+0.11}_{-0.12}$;
$\chi ^2 _\nu \sim$1.21 for 77 d.o.f.).  The observed 0.5--10~keV flux
of Mrk~273 is approximately 2.9 times that of Mrk~273x.  SBS~1342+560
is a background QSO at a redshift of 0.937 located approximately
6.3\arcmin\ south of Mrk~273.  The XIS spectrum of this source is also
well fit by a power-law component modified only by Galactic absorption
($\Gamma \sim$2.03$\pm 0.05$; $\chi ^2 _\nu \sim$1.28 for 242 d.o.f.).
The observed 0.5--10~keV flux of Mrk~273 is approximately 1.5 times
that of this source.  Lastly, SDSS~J1342512.06+554759.6 is a
background QSO at a redshift of 1.17 located approximately 7\arcmin\
southeast of Mrk~273.  Its XIS spectrum is once again well fit by a
power-law component modified only by Galactic absorption ($\Gamma
\sim$1.62$\pm0.07$; $\chi ^2 _\nu \sim$1.29 for 139 d.o.f.).  The
observed 0.5--10~keV flux of Mrk~273 is approximately 1.9 times that
of this source.

Thus, none of the objects within the XIS field of view shows evidence
of being highly obscured.  One would therefore expect their spectral
energy distributions at higher energies to follow the power law seen at
low energies.  If that is indeed the case, their contributions to the
overall PIN signal of Mrk~273 are negligible; this is explained in the
next section.

\subsubsection{Combined XIS-HXD/PIN Spectral Modeling}
\label{sec:m273fit}

\begin{figure}
\figurenum{3}
\epsscale{0.6}
\rotatebox{-90}{\plotone{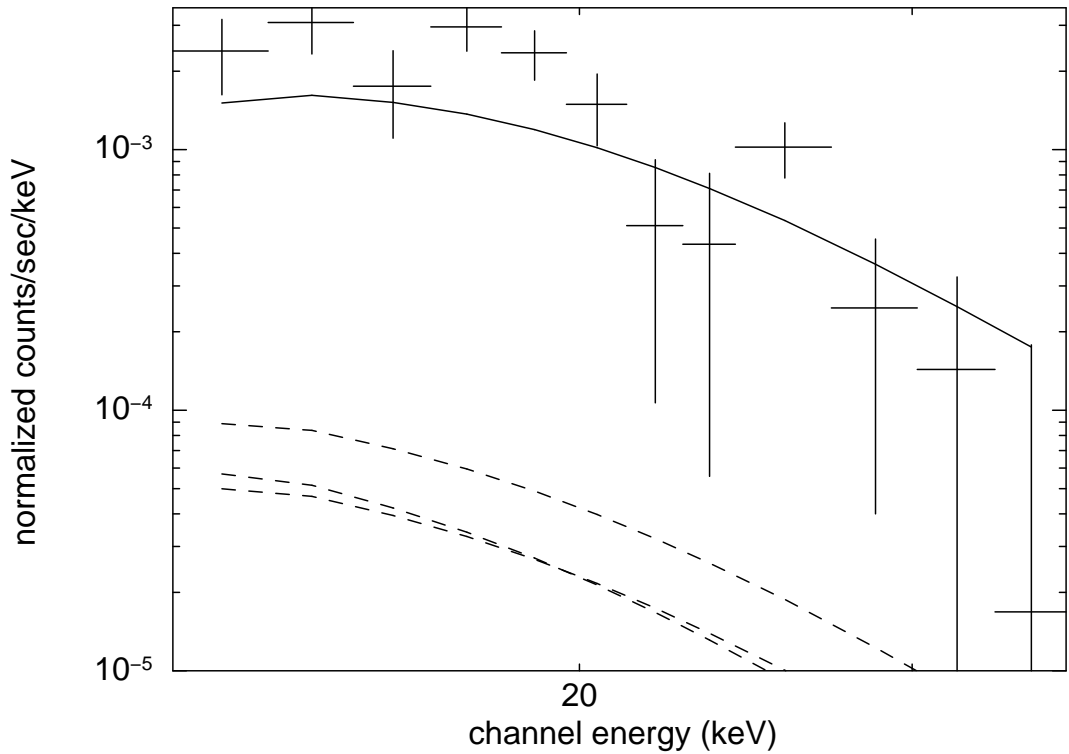}}
\caption{The net background-subtracted 15--40~keV HXD/PIN spectrum of
  Mrk~273 compared with the spectral models of possible contaminants in the PIN field of view.  The solid line represents the best-fit model from the
  XIS-HXD/PIN spectral fitting and is a sum of the flux from Mrk~273 and the contaminants.  The dashed lines represent
  the best-fit XIS models of the contaminants extrapolated to the
  HXD/PIN energy range.  The contributions from these contaminants to
  the overall PIN signals are negligible.}
\label{fig:hxdcontams}
\end{figure}

\begin{figure}
\figurenum{4}
\epsscale{0.6}
\rotatebox{-90}{\plotone{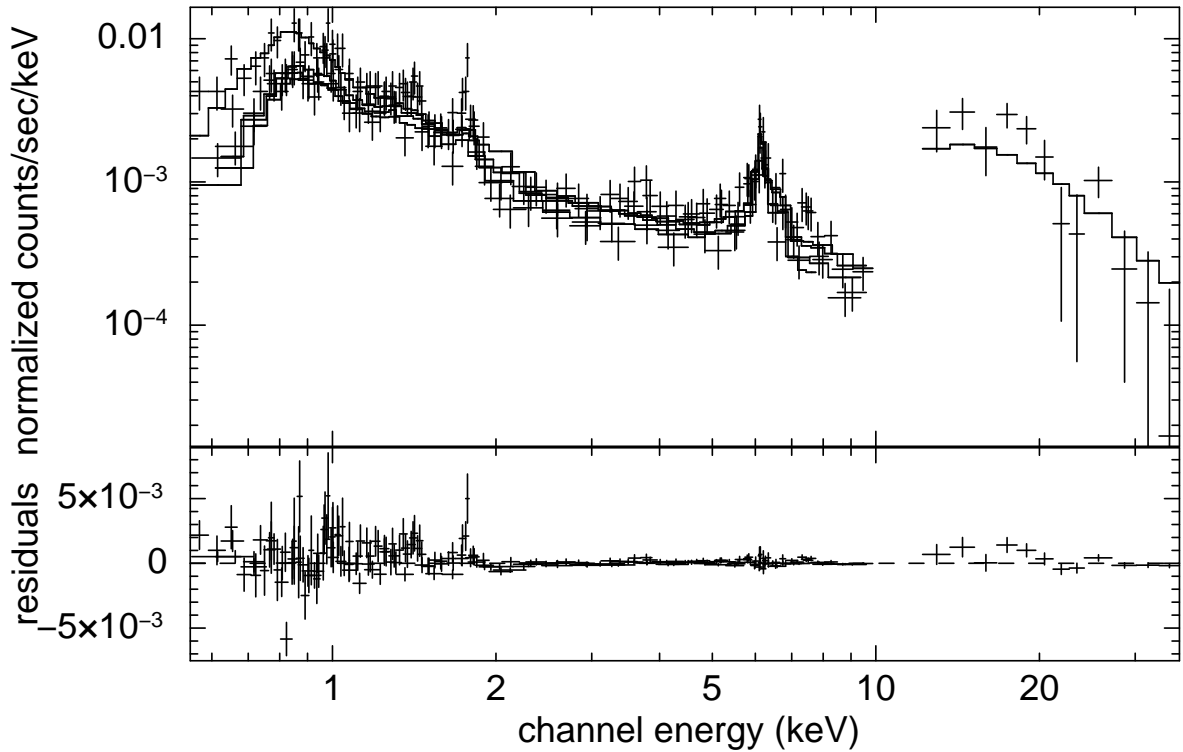}}
\caption{The combined XIS-HXD/PIN spectrum of Mrk~273 with the best-fit scattering model.  The
  horizontal axis is energy in the observer's frame. The directly
  transmitted AGN flux is $\sim$5\% of the intrinsic flux.
  The iron line at $\sim$6.4~keV is detected and a
  MEKAL model with gas temperature $\sim$0.7~keV is needed to reproduce
  the soft X-rays. The cross-normalization of the HXD/PIN with respect
  to XIS0 is assumed to be 1.13.  The best-fit parameters for the full band (XIS+HXD/PIN)
  modeling are listed in Table~\ref{tab:fithxd}.  }
\label{fig:hxdfig}
\end{figure}

We modeled the combined XIS and PIN spectrum of Mrk~273 with the
scattering model.  All of the best-fit parameters are
consistent with the values obtained from the fit of the XIS spectrum
alone with the exception of the column density which increased by
a factor of $\sim$2.  The HXD/PIN contributions from the neighboring
sources have been taken into account by adding the best-fit XIS model
of the contaminants to the HXD/PIN component of the Mrk~273 model (see
Figure~\ref{fig:hxdcontams}).  The best-fit model to the full-band
Mrk~273 spectrum in this case has parameter values listed in
Table~\ref{tab:fithxd} and the spectrum is shown in
Figure~\ref{fig:hxdfig} with the XIS-HXD/PIN spectrum.  While the photon index value of $\sim$1.4 may suggest flattening of the spectrum due to reflection, the addition of a reflection component resulted in a statistically worse fit ($\chi^2_\nu = 1.66$ for 197 d.o.f.) relative to that obtained from the scattering model ($\chi^2_\nu = 1.35$ for 199 d.o.f.).  Approximately 5\% of the intrinsic AGN
flux is scattered.   The gas
temperature of $\sim$0.7~keV is consistent with the range found in
ULIRGs \citep{grimes}.   Overall, the best-fit model of Mrk~273 agrees well with X-ray observations of
other ULIRGs \citep[e.g.][]{ptak, frances, teng, teng08}.  The comparison of the XIS data with published results is postponed
until \S~\ref{sec:dis273}.    

\subsection{PKS~1345+12}
\label{sec:pks1345}

\subsubsection{XIS Spectrum}
\label{sec:pks1345xis}

The XIS spectrum of PKS~1345+12 is shown in Figure~\ref{fig:xisfig}.
It is best modeled by a scattering continuum model
(\S~\ref{sec:results}) modified by Galactic absorption. The
transmitted AGN flux is $\sim$2\% of the intrinsic AGN flux.  Unlike
other ULIRGs, the 0.5--2~keV spectrum does not require a MEKAL thermal
component.  The best-fit parameters are listed in
Table~\ref{tab:fitxis}.  These best-fit values and the model of the
0.5--10~keV continuum are consistent with {\it Chandra} measurements
reported by \citet{iman04}.  The \citet{iman04} model for the {\it
Chandra} data includes a narrow Fe line at 6.4~keV.  This emission
line also appears in the XIS spectrum, but is statistically
insignificant ($\Delta \chi^2$=2.0 for a change in d.o.f.
of 2).  This is also consistent with the fact that the lower limit to
the equivalent width of the Fe line in the \citet{iman04} model is
0~keV.  No flux or spectral variation were detected
over the 0.5--10~keV energy range between the {\it Chandra} and {\it
Suzaku} observations.

The unabsorbed 0.5--10~keV flux for PKS~1345+12 is $\sim$6.0$\times
10^{43}$ ergs s$^{-1}$ and the absorption corrected ratio log[$L_{\rm
  2-10~keV}$/$L_{\rm IR}$] is $\sim$--2.3, consistent with the range
found in PG quasars.  This and the lack of a thermal component in 
the 0.5--2~keV spectrum suggest that the X-ray spectrum is dominated
by an AGN.

\subsubsection{HXD/PIN Data}
\label{sec:pks1345hxd}

The net HXD/PIN spectrum of PKS~1345+12 lies $\sim$3.4\% above the
total background (Figure~\ref{fig:speccomp}).  Accounting for the
uncertainties in the PIN background of $\sim$3.1\% (see
\S~\ref{sec:f05189hxd} on how this was calculated), the detection is
at only 1.1$\sigma$.  There are not enough net counts for a meaningful
spectral fitting of the full-band (XIS+HXD/PIN) spectrum.  However,
assuming no contamination from nearby sources, the net observed
15--40~keV count rate for PKS~1345+12 is 9.9$\pm$2.8$\times
10^{-3}$~cps.  Using the XIS best-fit model, given the lack of an
obvious reflection component, the 15--40~keV count rate translates to
a flux of $\sim$4.4$^{+1.1}_{-1.3} \times 10^{-12}$ ergs s$^{-1}$
cm$^{-2}$.

\subsection{Arp~220}
\label{sec:arp220}

\subsubsection{XIS Spectrum}
\label{sec:arp220xis}

Arp~220 is detected by the XIS. At first, the XIS spectrum was modeled by
a simple power-law model plus a MEKAL component for the thermal
contribution from the starburst.  An additional Gaussian component is needed to
reproduce the emission line at $\sim$6.7~keV (EW$\sim$1.98~keV), which is consistent with
emission from Fe~XX--Fe~XXVI.  This model with only Galactic absorption is a satisfactory fit to the XIS data ($\chi _\nu ^2 \sim$1.03 for 174 d.o.f.).  
The XIS extraction window encompasses the two nuclei of Arp~220 and a
significant fraction of the extended soft X-ray emission \citep{arp220b}.  Nevertheless, the power-law plus MEKAL model to the XIS-only data are
consistent with the results from the {\it Chandra} observation of the
nuclear binary source \citep{arp220a}.  There is, therefore, no obvious sign of spectral or flux variability in Arp~220.

The {\it Suzaku} data on Arp~220 can be interpreted in several ways:
(1) an AGN is not present, (2) there is a low-luminosity, unobscured
AGN, or (3) there is a heavily obscured AGN.  

In the first case where
an AGN is not present, the detected Fe line at $\sim$6.7~keV would
originate from hot gas heated by the heavy starburst activity.
However, the measured equivalent width for the line of
1.98$^{+0.99}_{-0.57}$~keV seems too large for an Fe emission line
arising from a pure starburst.  
\citet{iwasawa} fitted the 2.5--10~keV continuum of the {\it XMM-Newton} data with
a collisionally ionized plasma model.  The model requires a gas
temperature of $\sim$7~keV and a metallicity of $\sim$2 times solar in
order to reproduce the 6.7~keV Fe feature (EW$\sim$2~keV).  
This temperature is inconsistent with the detected Ca XIX line in the
{\it XMM-Newton} data.  The XIS data do suggest the presence of an
emission feature at $\sim$3.9 keV consistent with Ca XIX, but this
detection is not statistically significant.

The weak but unobscured AGN case is unlikely based on CO observations
by \citet{downes07}.  Their observations with IRAM imply a column
density of $\sim 1.3 \times 10^{25}$ cm$^{-2}$ for the western nucleus
of Arp~220.  Unless the covering fraction of the absorber is much less
than unity, the unobscured AGN interpretation of the data is
inconsistent with the CO observations.  

In contrast, the third scenario of a heavily obscured AGN is
consistent with the Fe line detection and the CO observation of a
large column density.  The ionized reflection model as proposed by
\citet[][and references therein]{reflion} can reconcile the large
equivalent width of the highly ionized iron line as well as the large
absorbing column.  Since the absorbing column is so large, the
observed spectrum is purely reflected.  The XIS data is best fit by a
MEKAL plus an ionized reflection model with the ionization parameter
($\xi$) fixed at 10$^3$ ergs cm s$^{-1}$ ($\chi^2_\nu = 1.03$ for 176
d.o.f.; see Table~\ref{tab:fitxis} and Figure~\ref{fig:xisfig}).  Thus
the third scenario for explaining the {\it Suzaku} data on Arp~220 is
favored.

Based solely on the XIS data, the unabsorbed 0.5--10~keV luminosity of
Arp~220 is only $\sim 1.8 \times 10^{41}$ ergs s$^{-1}$ cm$^{-2}$,
much lower than that found in quasars.  The absorption-corrected ratio
log[$L_{\rm 2-10~keV}$/$L_{\rm IR}$] is $\sim$--4.9, consistent with
those found in some nearby ULIRGs \citep{teng}, but lower than those
of radio-quiet quasars.  Of course, if Arp220 is highly absorbed then
the intrinsic log[$L_{\rm 2-10~keV}$/$L_{\rm IR}$] is higher and may
be within the range for PG quasars.  The thermal component of the
spectrum has a 0.5--2~keV luminosity of $\sim 3 \times 10^{40}$ ergs
s$^{-1}$.  If this luminosity is completely due to thermal
bremsstrahlung, then applying this to Equation (1) of \citet{teng08},
the spatial extent of the emitting region is $\sim 2-11$ kpc, depending
on the choice of filling factor ($10^{-3} < f_{gas} < 10^{-1}$) and assuming an electron density of 1 cm$^{-3}$.

\subsubsection{HXD/PIN Data}
\label{sec:arp220hxd}

As shown in Figure~\ref{fig:speccomp}, the net HXD/PIN spectrum of
Arp~220 is only $\sim$2.2\% above the (NXB+CXB) background , so it is
within the uncertainties of the PIN background modeling ($\sim$3.0\%,
calculated following procedures in \S~\ref{sec:f05189hxd}).  Thus, the
estimated 15--40~keV count rate for Arp~220 is $\sim$0.025 cps.

\section{The Long-term Variability of {\em IRAS}~$F$05189--2524 and Mrk~273}
\label{sec:var}

Within the span of the {\it Suzaku} observations ($\sim$80~ksec each),
no significant variability is detected in the XIS data for both
$F$05189--2524 and Mrk~273.  However, there is evidence for long-term
variability in both sources when the {\it Suzaku} data are compared
with previous {\it ASCA, BeppoSax, Chandra,} and {\em XMM-Newton}
data. In this section, we discuss the results of these comparisons and
their implications for the spectral models. 

\subsection{{\em IRAS}~$F$05189--2524}
\label{sec:dis05189}

\begin{figure}
\figurenum{5}
\epsscale{1}
\plotone{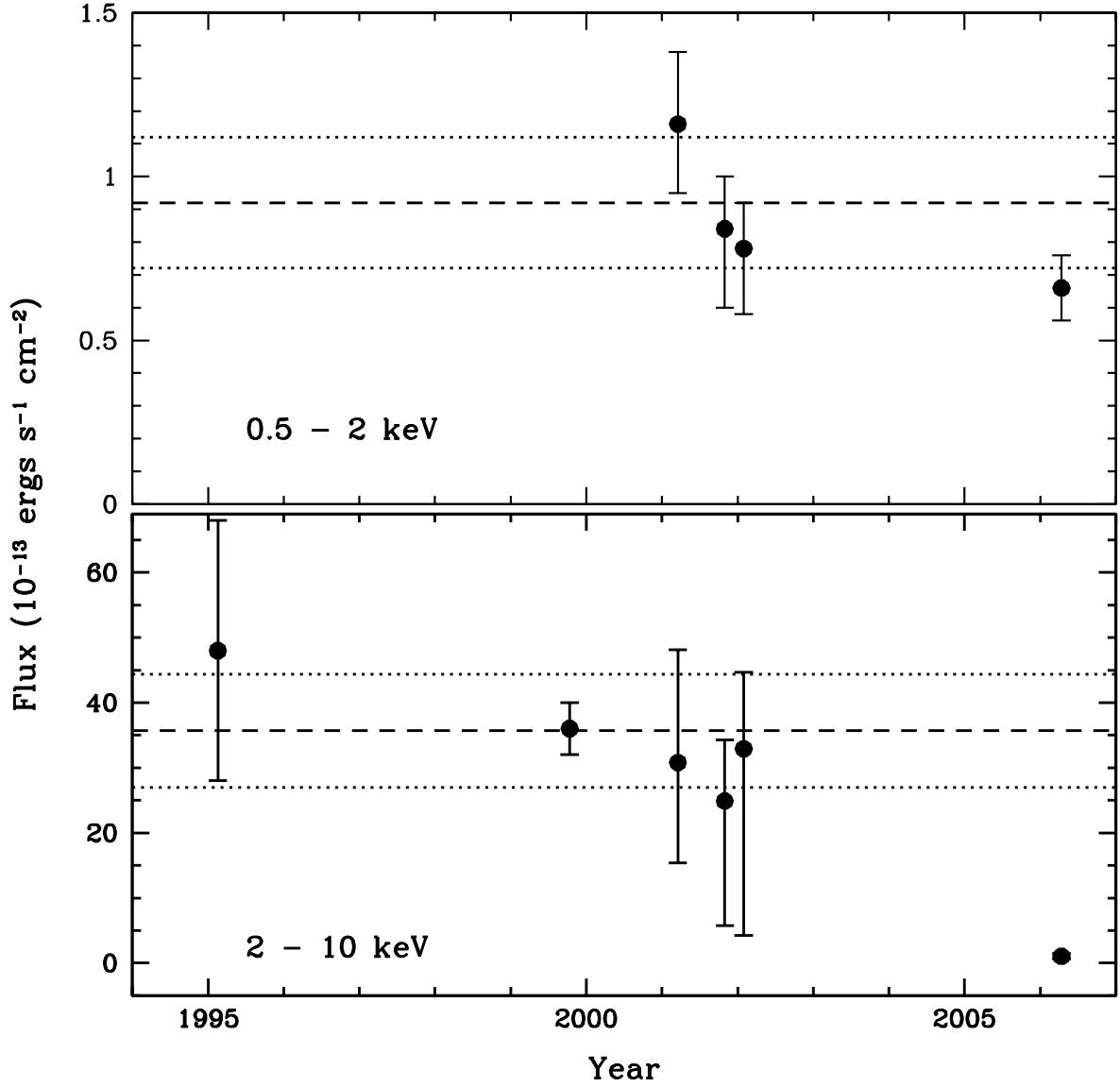}
\caption{Observed 0.5--2~keV (top) and 2--10~keV (bottom) fluxes of
  $F$05189--2524 from 1995 to 2006 as
  determined from {\it ASCA}, {\it BeppoSAX}, {\it XMM-Newton}, {\it
  Chandra}, and {\it Suzaku} data.  While there has been little change
in the 0.5--2~keV flux, the nominal 2--10~keV flux of $F$05189--2524 has 
decreased by a factor of $\sim$30 since previous observations.  The dashed line represents the average ``high'' state flux (measurements made prior to {\it Suzaku}) as weighted by the measurement errors while the dotted lines denote one standard deviation away from the mean.  The 2--10~keV flux value as measured by {\it Suzaku} is $\sim$4 standard deviations away from the weighted mean.  The
{\it ASCA} (1995) and {\it BeppoSAX} (1999) values are drawn from
\citet{severgnini}; the {\it XMM-Newton} (2001) and {\it Chandra} (2002)
values are drawn from our modeling of the archived spectra with
the scattering model in this work (see \S\ref{sec:dis05189}).  The error bars for the 2006 value are within the data point.}
\label{fig:05189var}
\end{figure}

\begin{figure}
\figurenum{6}
\epsscale{.8}
\rotatebox{-90}{\plotone{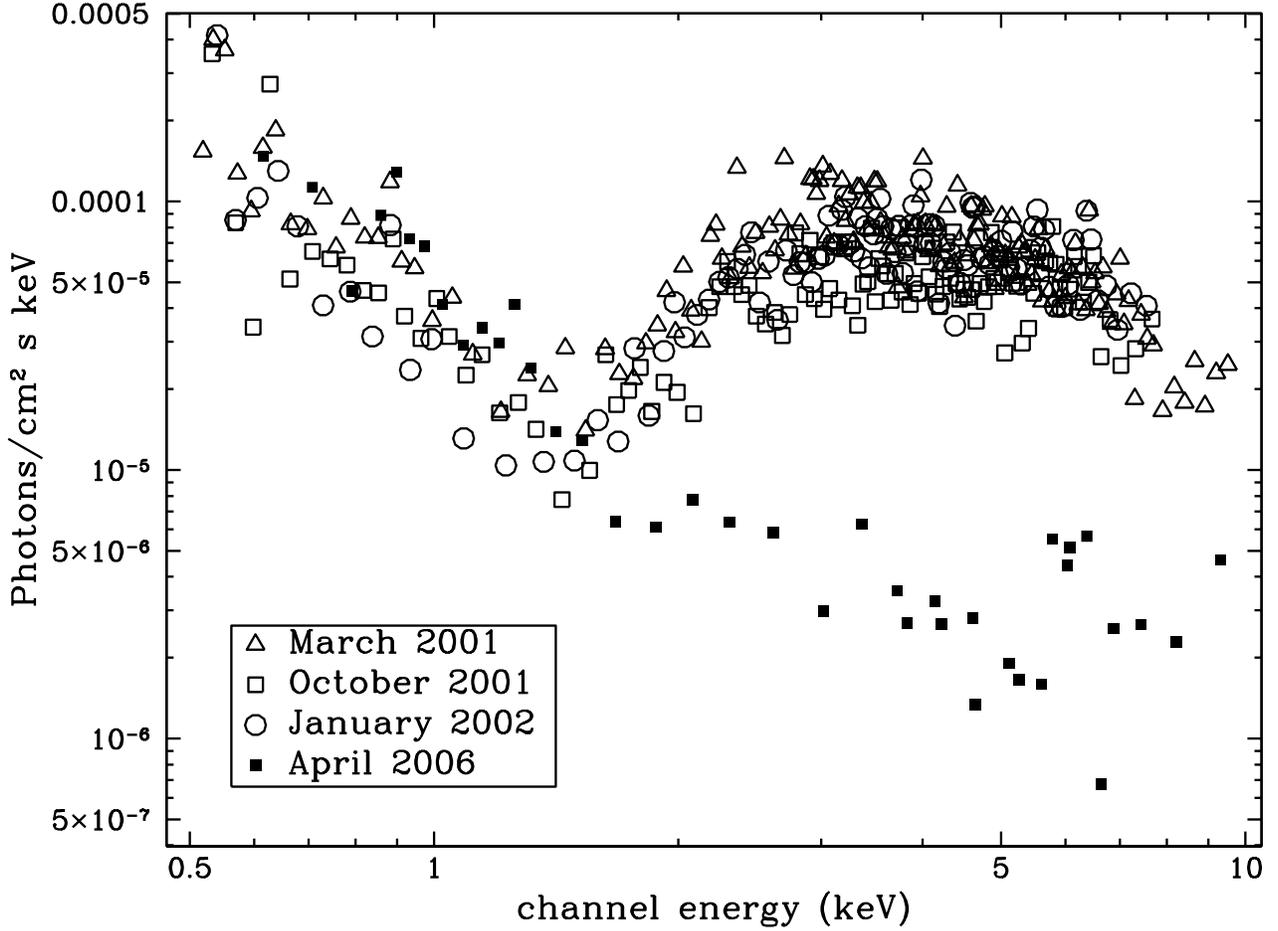}}
\caption{Comparison of the unfolded spectra of $F$05189--2524 from the 2001 March {\it XMM-Newton} (open triangles), 2001 October {\it Chandra} (open squares), 2002 January {\it Chandra} (open circles), and 2006 April {\it Suzaku} (closed squares) observations.  The unfolded spectra were created using the best-fit models to each individual spectrum.  Below $\sim$1.3~keV, there appears to have been negligible change in flux or spectral shape between the different observations.  However, above $\sim$1.3~keV, the spectral change is obvious.  The iron line is prominent in the {\it Suzaku} data, but is not noticeable in the other observations.  The best-fit models to the spectrum changed from a scattering-dominated scenario in 2001 and 2002 to a reflection-dominated scenario in 2006.  This may be an indication that the central source has faded prior to 2006 or variations in the column density.}

\label{fig:05189uf}
\end{figure}

$F$05189--2524 was observed by {\em XMM-Newton} in March 2001 and then
by {\it Chandra} in October 2001 and January 2002.  We extracted the
spectra from the {\em XMM-Newton} and {\it Chandra} archives.
There do not appear to be significant variations in the 0.5--2~keV flux
of $F$05189--2524.  The 2006 {\it Suzaku} measurement is within a
standard deviation of the weighted average of the previous
measurements.  In contrast, the {\it Suzaku} 2--10~keV flux is a
factor of $\sim$30 lower than the measurements made by other
observatories, $\sim$4 standard deviations away from the weighted
average of the previous measurements (see Figure~\ref{fig:05189var}
and \ref{fig:05189uf}).

None of the {\it XMM-Newton} and {\it Chandra} data shows a
significant iron feature near 6.4~keV, in contrast to our more recent
{\em Suzaku} data, so we did not consider reflection models for these
archived data.  The scattering model (\S~\ref{sec:results}) with a
thermal MEKAL component gave adequate fits to the spectra.  In the
modeling of these data, we allowed the internal column density and the
parameters for the power-law component describing the AGN within the
source to vary freely.  The models suggest that the archived {\em
XMM-Newton} and the two sets of {\em Chandra} spectra are absorbed by
a column density of 7.6$\times 10^{22}$, 6.8$\times 10^{22}$, and
6.1$\times 10^{22}$ cm$^{-2}$, respectively.  We also deduce that the
observed 2--10~keV flux of the AGN has decreased from 3$\times
10^{-12}$ ergs s$^{-1}$ cm$^{-2}$ in the 2001 {\em XMM-Newton} data to
1$\times 10^{-13}$ ergs s$^{-1}$ cm$^{-2}$ in the 2006 {\it Suzaku}
data {\em i.e.} a reduction of a factor of $\sim$30.

The decrease in flux and the change in spectral shape may be due to a
number of reasons.  Here, we test three scenarios by modeling the
multiple epoch data simultaneously with a single model.  The three
scenarios considered are: (1) a change in the column density while the
properties of the AGN remain the same, (2) a
change in the covering fraction of the absorber while the properties
of the AGN remain the same, and (3) the column density and covering
fraction remain the same while the intrinsic AGN luminosity is
changed.  In this scenario, a reflection component is visible in the ``low-state'' AGN.

\begin{figure}
\figurenum{7}
\epsscale{0.7}
\plotone{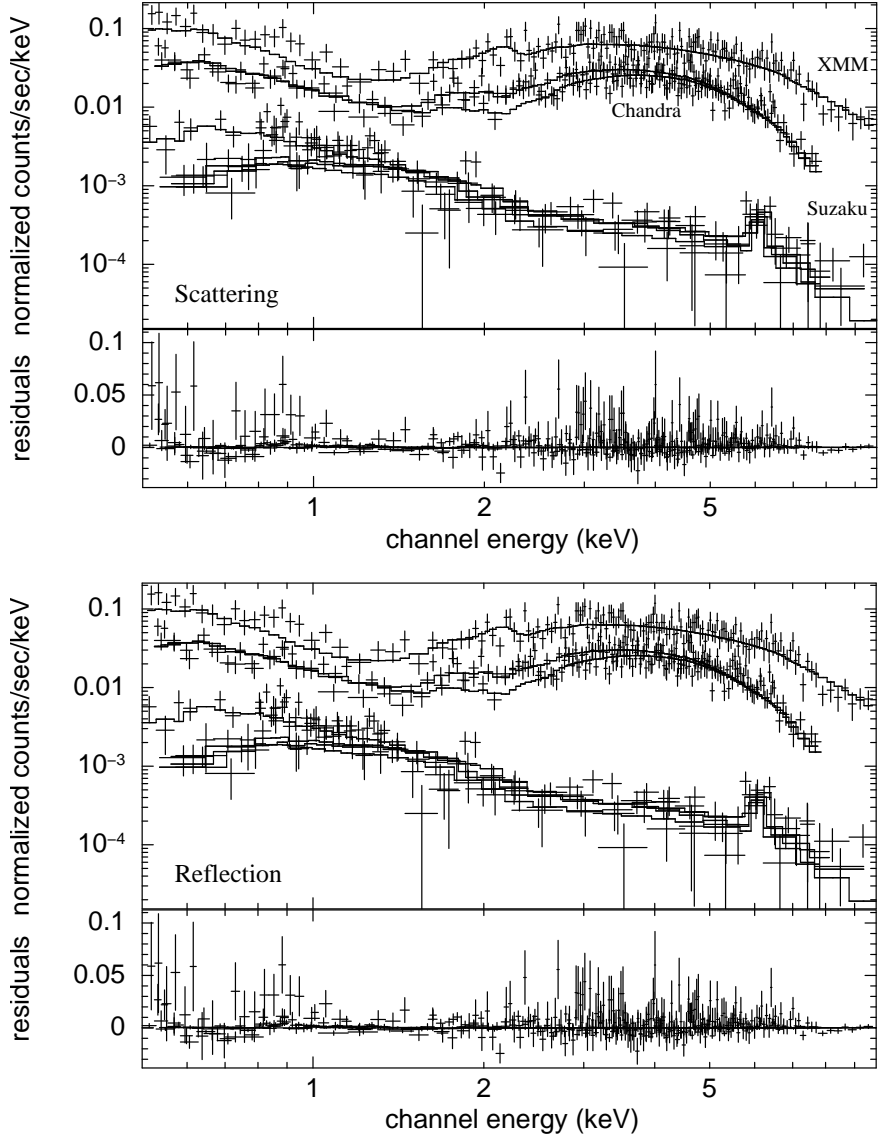}
\caption{The two models to the {\em Chandra}-ACIS, {\em
    XMM}-Newton-EPIC pn, and {\it Suzaku}-XIS/PIN data used to explain
  the change in the 2--10~keV spectral shape of $F$05189--2524:
  (top) change in the absorbing column alters the spectral shape of the source, and (bottom) the AGN has switched off and has left behind a residual reflection component (see \S\ref{sec:dis05189} for details).  The response
  from each detector is folded in with the data.  The horizontal axis
  is energy in the observer's frame. The cross-normalization factor is
  assumed to be unity for {\em XMM-Newton} and {\it Chandra} with
  respect to {\it Suzaku}-XIS0.  Both models have comparable
  reduced $\chi ^2$ values (1.27 for 503 d.o.f. versus 1.30 for 502 d.o.f.) and appear to fit the
  overall 0.5--10~keV spectrum well.  Neither model can be ruled out by the data.  Table~\ref{tab:fitmulti} lists the results from this modeling.}
\label{fig:05189multi}
\end{figure}

The first scenario is tested by modeling all four data sets
simultaneously by a single scattering model plus a MEKAL component
modified only by Galactic absorption and a Gaussian describing the
iron line.  All of the model parameters are set to be the same for
each of the data groups, except for the column density parameter which
is allowed to vary independently.  This results in a satisfactory fit
to the data ($\chi ^{2}_\nu$ = 1.26 for 503 d.o.f.).  This model and
the multi-epoch data are shown in Figure~\ref{fig:05189multi} and the
best-fit parameters are listed in Table~\ref{tab:fitmulti}.

For the second scenario, the multi-epoch data are simultaneously fit
by a single partial covering fraction model.  All of the model
parameters are set to be the same between epochs, except for the
covering fraction.  The modeling of this scenario results in an
unacceptable fit with $\chi ^2_\nu \sim 8$.  The addition of another
partial-covering absorber to this model results in a still
unsatisfactory fit with $\chi^2_\nu = 2.2$ for 498 degrees of freedom.
Thus the partial-covering model is not a good description of the data.

Lastly, the third scenario is examined.  In this scenario, the active
nucleus has switched off prior to the final set of observations,
leaving behind a residual reflection component visible in the ``low''
state.  This has been observed previously in NGC~4051 \citep{n4051} and NGC~1365 \citep{n1365}.  The model for
this scenario is a combination of a MEKAL component, plus a reflection
component on top of a scattering model.  A Gaussian is also added to
model the Fe line seen in the {\it Suzaku} data.  For the {\it
Suzaku}-epoch, the scattered component normalization is held fixed at
zero.  Similarly, for the non-{\it Suzaku}-epochs, the reflected
component normalization is fixed at zero.  Additionally, the
absorption parameter in each of the non-{\it Suzaku}-epochs is
allowed to vary independently while the {\it Suzaku}-epoch absorption
parameter is held fixed at zero.  Since the intrinsic
AGN luminosity has changed, the normalization factors for the direct
component are different for the {\it Suzaku}- and non-{\it
Suzaku}-epochs.  The modeling of the data with this scenario is rather
successful with $\chi^2_\nu = 1.30$ for 502 degrees of freedom.  The
best-fit parameters to this model are listed in
Table~\ref{tab:fitmulti} and the model shown in
Figure~\ref{fig:05189multi}.

Both the increase in column density and the decrease in intrinsic
luminosity of the AGN appear to be equally good descriptions of the
data.  However, in the first scenario, the $N_{\rm H}$ value required
to explain the data is $> 2 \times 10^{24}$ cm$^{-2}$.  At this limit,
we can no longer model it as a pure
absorption spectrum since scattering both in and out of the line of sight are becoming
important. 

The thermal 0.5--2~keV luminosity is $\sim 1 \times 10^{41}$ ergs
s$^{-1}$.  Following the methods presented in \S~\ref{sec:arp220xis},
the physical extent of the emitting region is $\sim$4--17~kpc).  The absorption-corrected 0.5--10~keV
luminosity of the AGN detected in $F$05189--2524 at its ``high'' state (2000--2002 observations)
is $\sim$3$\times 10^{43}$ ergs s$^{-1}$, almost as luminous as some
quasars \citep{elvis}.  The absorption-corrected ratio log[$L_{\rm
2-10~keV}$/$L_{\rm IR}$] is $\sim$--2.5, consistent with that found in
radio-quiet PG quasars.  The corresponding numbers for the ``low'' state as observed by {\it Suzaku} are $\sim$8$\times 10^{41}$ ergs s$^{-1}$ and --4.1, respectively, well below that of radio-quiet PG quasars.

\subsection{Mrk~273}
\label{sec:dis273}

\begin{figure}
\figurenum{8}
\epsscale{.8}
\plotone{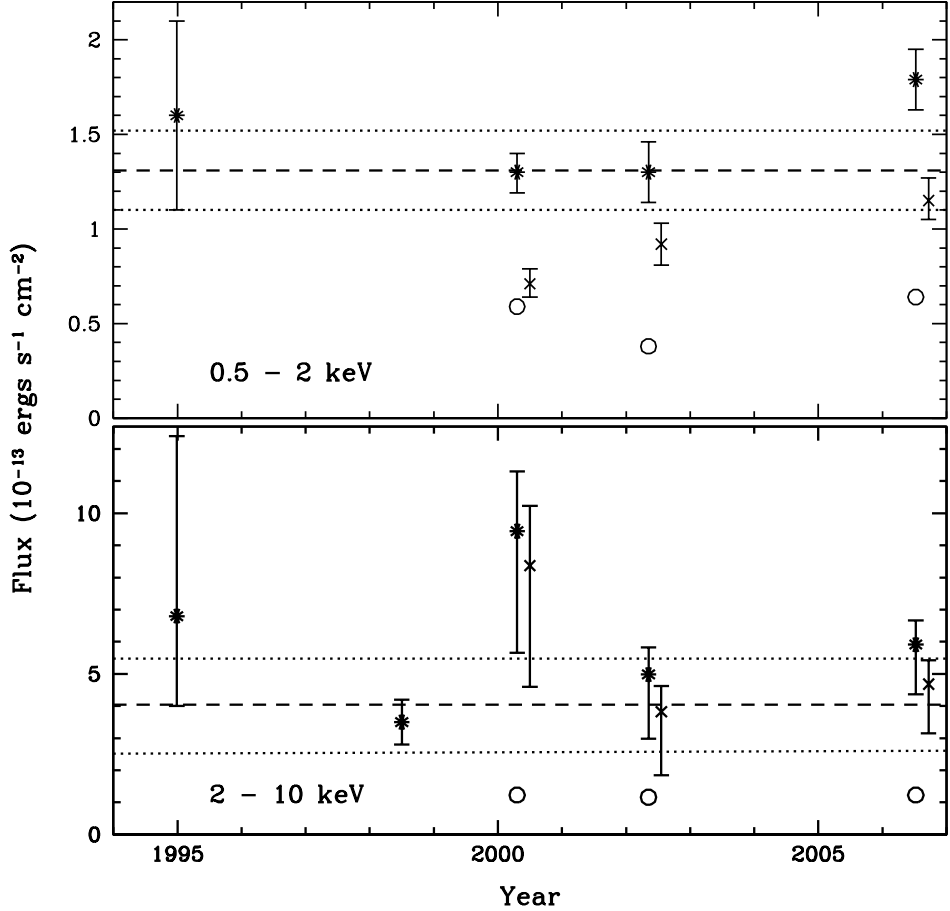}
\caption{Observed 0.5--2~keV (top) and 2--10~keV (bottom) fluxes of
  Mrk~273 (crosses) and Mrk~273x (open circles) from 1994 to 2006 as
  determined from {\it ASCA}, {\it BeppoSAX}, {\it XMM-Newton}, {\it
  Chandra}, and {\it Suzaku} data.  Each asterisk represents the sum of
  the fluxes from Mrk~273 and Mrk~273x.  In the top panel, the dashed line represents the weighted average of the data points for Mrk~273 and Mrk~273x from 1994, 2000, and 2002 and the dotted lines represent a single standard deviation away from that mean.  In the bottom panel, the dashed line represents the weighted average of the data points for Mrk~273 and Mrk~273x from 1994, 1998, 2002, and 2006 and the dotted lines denote one standard deviation away from the mean.  Small flux increases
  in Mrk~273 may be observed in 2006 in the 0.5--2~keV band and in 2000 in the 2--10~keV band. 
The {\it ASCA} (1994) and {\it BeppoSAX} (1998)  values are drawn from
  \citet{iwasawa99} and \citet{risaliti00}, respectively.  The {\it
    Chandra} (2000) and {\it XMM-Newton} (2002) values are derived
  from our modeling of the archived spectra with the
  scattering model (see \S\ref{sec:dis273}).  The error bars to the total flux
  are the sums of the errors in the measured fluxes of Mrk~273 and
  Mrk~273x added in quadrature.  A small horizontal offset was applied to the Mrk~273 data points to better display the measurement errors.  Since the errors on the {\it
    BeppoSAX} measurements are not published, they are assumed to be
  $\pm$20\% based on the results on $F05189-2524$ by
  \citet{severgnini}.}
\label{fig:273var}
\end{figure}

\citet{xia} compiled a list of observed fluxes of Mrk~273 and Mrk~273x
from {\it ASCA, BeppoSax,} and {\it Chandra} dating from 1994 to
2000\footnote{Although the authors tabulated the fluxes from 1996 to
  2000, the {\it ASCA} observation date was mis-identified and the
  data set was actually taken in December of
  1994 \citep[see][]{iwasawa99}.}.
We add to this list the {\em XMM-Newton} observation from 2002
analyzed by \citet{balestra} and our {\it Suzaku} observation from
2006.  
Figure~\ref{fig:273var} shows the observed fluxes of Mrk~273 and Mrk~273x over the 12-year period.  Since {\it
  ASCA} and {\it BeppoSax} did not have the spatial resolution to
separate Mrk~273 from Mrk~273x, the flux points from 1994 and 1998
represent the sum of the two galaxies.  As the figure shows, the total flux between Mrk~273 and Mrk~273x appears to have increased in 2006 in the 0.5--2~keV band and appears to be in a ``high'' flux state in 2000 relative to all the other measurements.

Although Mrk~273x was once thought to be a BL Lacertae object because
of its high X-ray to optical B-band flux ratio \citep{xia98}, its
2--10~keV flux has never shown dramatic variability.  The {\it Suzaku}
data support the assessment of \citet{xia} that Mrk~273x is
unlikely to be a BL Lac object.  As discussed in \S 4.3.2, its
spectrum is well fit by a simple power law modified only by Galactic
absorption, in agreement with the findings in \citet{balestra}.

\begin{figure}
\figurenum{9}
\epsscale{.8}
\rotatebox{-90}{\plotone{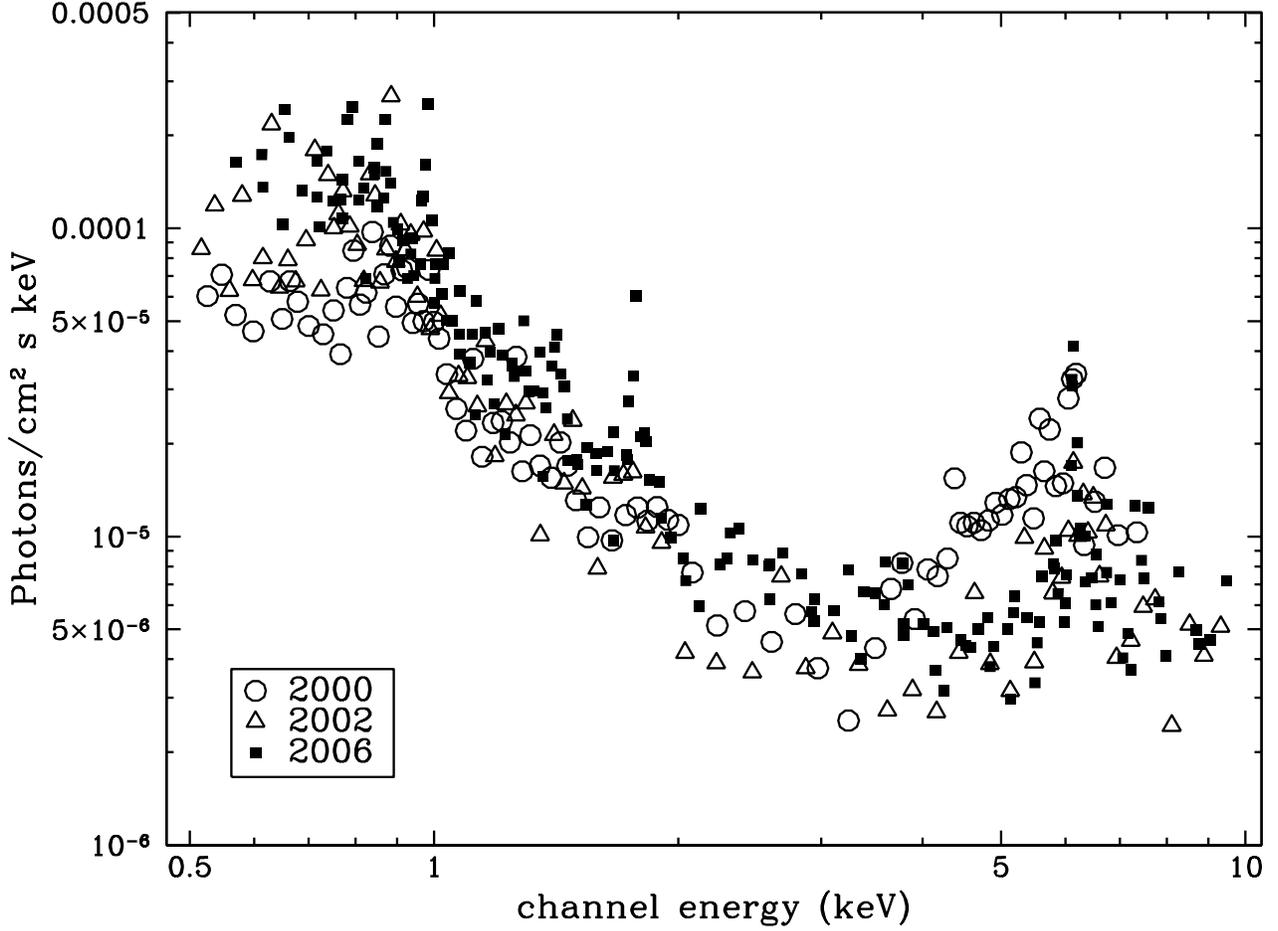}}
\caption{Comparison of the unfolded spectra of Mrk 273 from the 2000
{\it Chandra} (open circles), 2002 {\it XMM-Newton} (open triangles),
and 2006 {\it Suzaku} (closed squares) observations.  The unfolded
spectra were created using the best-fit models to each individual
spectrum.  The shapes of the spectra appear consistent between 1 and
2~keV.  The flux of the {\it Suzaku} spectrum is higher than that of
the others below 1~keV, while the {\it Chandra} spectrum appears to be
higher than the others between 4 and 6~keV.  These differences are
correlated with the flux variability as shown in
Figure~\ref{fig:273var}.  This comparison also shows that the iron
line at 6.4~keV is detected in all three observations.  While the
2--10~keV flux variability is not very significant, the spectral shape
of the source has changed between 2000 and 2006.  This change may be
due to the variations in the column density.}
\label{fig:m273uf}
\end{figure}

Assuming a constant flux for Mrk~273x implies that the nominal observed
2--10~keV flux from Mrk~273 dropped by about a factor of two from
1994 to 1998, then increased by a factor of $\sim$2.5 from 1998 to
2000, and then again dropped by a factor of more than two from 2000 to
2002. It appears to have remained roughly the same from 2002 to 2006.  This variability in flux is not very significant when considering the uncertainties of the measurements.  However, the spectral variability is clearly seen when comparing the spectra of Mrk~273 from {\it Chandra}, {\it XMM-Newton}, and {\it Suzaku} (Figure \ref{fig:m273uf}).  The higher 0.5--2~keV flux in 2006 and the higher 2--10~keV flux in 2000 as shown in Figure~\ref{fig:273var} correlate with the difference in the spectral shapes of the source in each of these epochs.  

\subsubsection{Modeling the Multiple Epoch Data}
\label{sec:273epochs}

As with the multiple data sets of $F$05189--2524, we first modeled the
archived {\it Chandra} and {\em XMM-Newton} spectra of Mrk~273
individually. These data sets were each best fit by the scattering model.  The fits to the {\it Chandra} ({\em XMM-Newton})
data imply an intrinsic column density of 4.1 (6.8) $\times 10^{23}$
cm$^{-2}$ and an intrinsic 2--10~keV flux of 2.5 (2.0) $\times
10^{-12}$ ergs s$^{-2}$ cm$^{-2}$. A comparison of the {\it Chandra}
({\it XMM-Newton}) data with the fits to the
2006 {\it Suzaku} XIS data in Table~\ref{tab:fitxis} suggests that the intrinsic column density has
increased by $\sim$110\% ($\sim$27\%) while the intrinsic 2--10~keV luminosity has
decreased by $\sim$80\% (0\%) if we were to compare data only over the
0.5--10~keV range.  However, the addition of the HXD/PIN data
(Table~\ref{tab:fithxd}) implies
that the column density has increased by a factor of $\sim$4 ($\sim$2) and the
intrinsic 2--10~keV luminosity has increased by a factor of $\sim$3 ($\sim$4).

Next, all three data sets are modeled with a
single model simultaneously to determine whether the change in the
spectral shape is due to the change in column density or the intrinsic
AGN luminosity.  Three scenarios are considered: the change in
spectral shape is
due to (1) a change in the absorbing column and the intrinsic AGN
flux remains the same, (2) a
change in the intrinsic AGN luminosity with the absorbing column
remaining constant, and (3) a change in the covering fraction of the absorber with the column density and intrinsic
luminosity remaining constant.

\begin{figure}
\figurenum{10}
\epsscale{0.5}
\plotone{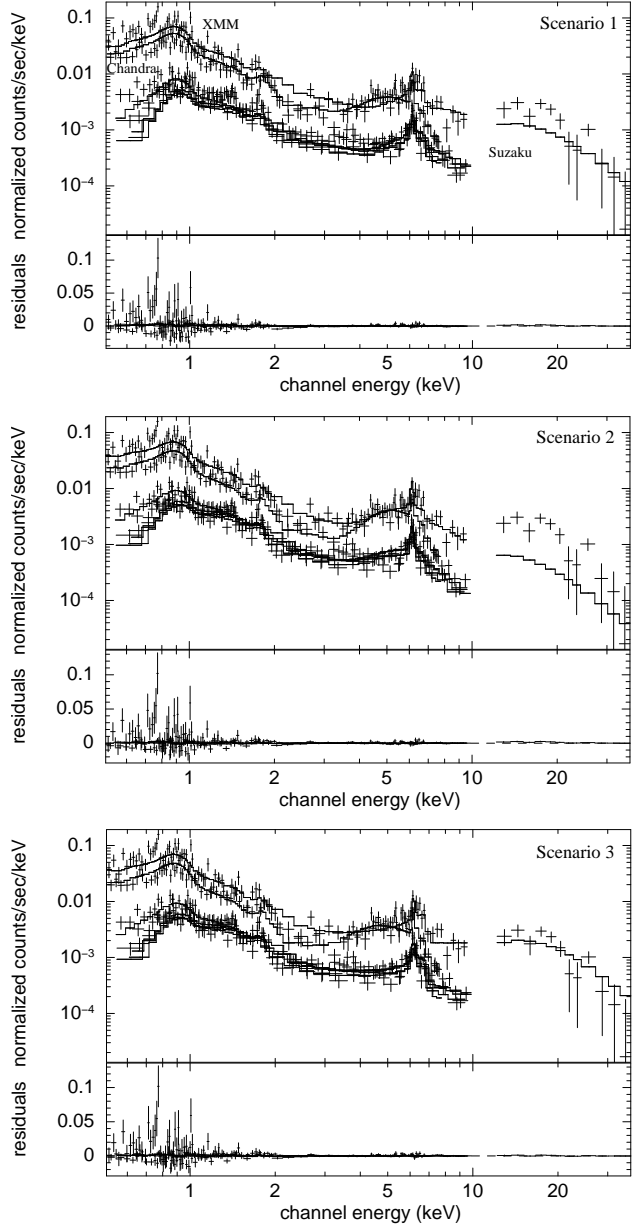}
\caption{The three models to the {\em Chandra}-ACIS, {\em
    XMM}-Newton-EPIC pn, and {\it Suzaku}-XIS/PIN data used to explain
  the changes in the 2--10~keV spectral shape of Mrk~273: scenario 1
  (top) tests only a change in the absorbing column, scenario 2
  (middle) tests only a change in the intrinsic luminosity of the AGN,
  and scenario 3 (bottom) tests the change in the covering fraction of
  the absorber (see \S\ref{sec:dis273} for details).  The response
  from each detector is folded in with the data.  The horizontal axis
  is energy in the observer's frame. The cross-normalization factor is
  assumed to be unity for {\em XMM-Newton} and {\it Chandra} with
  respect to {\it Suzaku}-XIS0.  While all three models have similar
  reduced $\chi ^2$ values (1.76, 1.53, and 1.39 respectively), the
  third model provides the best-fit to the overall 0.5--40~keV
  spectrum.  Thus, we favor the third scenario (change in the covering
  fraction of the absorbers) as the best explanation for the change in
  the 2--10~keV spectral shape of Mrk~273.  Table~\ref{tab:fitmulti}
  lists the results from this modeling.}
\label{fig:273multi}
\end{figure}

To evaluate the first scenario, all three data sets are
simultaneously modeled by a scattering plus MEKAL model
modified only by Galactic absorption and a Gaussian describing the
iron line.  All of the model parameters
are set to be the same for all of the data groups, but the column density
for data sets from each epoch is allowed to vary (assuming the intrinsic luminosity
remains the same).  As before, the contaminants in the HXD/PIN field
of view are taken into account in the modeling.  This interpretation of
the data results in a poor fit ($\chi^2_\nu = 1.76$ for 380 d.o.f.), much worse than the modeling to each data set alone and underestimates the flux of the HXD/PIN data (see Figure~\ref{fig:273multi}).  

The second scenario is explored by again modeling the multi-epoch data simultaneously
with a single model.  The model is the same as that used in the first
scenario.  However, in this case, the model parameters for each epoch are set to be the same except
for the normalization of the power law.  In this case, we are testing
whether the change in the intensity of the power-law component can
describe the observations.  The best-fit model for this scenario takes
into account the contributions from the nearby contaminants.  While the model describes the
0.5--10~keV data for each epoch well ($\chi^2_\nu = 1.53$ for 378 d.o.f.), it fits the HXD/PIN data very
poorly, severely under-estimating the HXD/PIN flux (see Figure~\ref{fig:273multi}).  

To test the third scenario, the {\it Chandra}, {\it
  XMM-Newton}, and {\it Suzaku} spectra are modeled using two absorbers with the
  same column densities for data from each epoch but with covering fractions that are allowed
  to vary freely. The contributions to the HXD/PIN flux from the
  neighboring XIS sources are taken into account in the fitting of
  the {\em Suzaku} data.  The best-fit model ($\chi^2_\nu = 1.39$ for 376 d.o.f.) is shown in
  Figure~\ref{fig:273multi} and the best-fit parameter values are
  listed in Table~\ref{tab:fitmulti}.  Based on Figure~\ref{fig:273multi}, the third scenario is the preferred
  model for the multiple epoch data.

Using the parameters derived for this third scenario, the
absorption-corrected luminosity from the AGN is $4.0 \times
10^{43}$~ergs~s$^{-1}$ in the 0.5--10~keV band and $2.3 \times
10^{43}$~ergs~s$^{-1}$ in the 15--40~keV band.  The ratio log[$L_{\rm 2-10~keV}$/$L_{\rm IR}$] is $\sim$--2.3,
within the range found in radio-quiet PG quasars.  The
thermal 0.5--2~keV luminosity is $\sim 1 \times 10^{40}$ ergs
s$^{-1}$.  Following the methods discussed in \S~\ref{sec:arp220xis},the emission region is $\sim$4--17~kpc in
size. 

Given the lack of knowledge of the detailed geometry of the nuclear region, it is difficult to derive a quantitative measure of the distance from the central source to the absorber.  However, the year-to-year time variability suggests that the absorber is close ($\sim$1~pc) to the nucleus.

\subsubsection{Comparison with OSSE Data}
\label{sec:273comp}

\citet{dermer} observed Mrk~273 with the OSSE instrument on-board the
{\it Compton Gamma Ray Observatory}.  Though their observations
resulted in a non-detection, they placed upper limits on the
50--100~keV band flux and the column density.  Possible explanations suggested by
\citet{dermer} for the non-detection include the possibility that this ULIRG is gamma-ray weak or the AGN
source is highly variable in gamma-rays and they happened to have
observed the object in the low-flux state.  

Another explanation for
the non-detection is that the AGN in Mrk~273 is hidden behind a large and patchy
column of gas $\gtrsim 10^{24}$~cm$^{-2}$.  This is not inconsistent
with the upper limit derived from previous CO measurements ($\lesssim
2 \times 10^{24}$~cm$^{-2}$, assuming a covering fraction of unity;
see Dermer et al. 1997 for more detail).  These limits on the column
density are consistent with our partial covering absorption model
($N_{\rm H} \sim 1.6 \times 10^{24}$ cm$^{-2}$).  From an
extrapolation of the HXD/PIN observation and model, we derive an upper
limit to the 50--100~keV photon flux of $6.2 \times
10^{-5}$~photons~cm$^{-2}$ or $6.9 \times
10^{-12}$~ergs~s$^{-1}$~cm$^{-2}$.  This value is more stringent by a
factor of $\sim$2 than the OSSE value ($1.5 \times
10^{-4}$~photons~cm$^{-2}$).

\section{Summary}
\label{sec:summary}

The results of our analysis of {\it Suzaku} XIS (0.5--10~keV) and
HXD/PIN (15--40~keV) observations of five of the brightest and
best-known local ULIRGs ($F$05189--2524, $F$08572+3915, Mrk~273,
PKS~1345+12, and Arp~220)
have been presented and compared with earlier {\em Chandra} and {\em
  XMM}-Newton data. The results can be summarized as follows:

\begin{enumerate}

\item The XIS observations of $F$05189--2524 reveal a significant change in the observed 2--10~keV
  spectrum relative to previous {\it Chandra} and {\em XMM-Newton}
  observations.  The spectral
  variation in $F$05189--2524 suggests that the central source may have turned off, leaving behind a residual reflection component.  However, an increase in column density cannot be completely ruled out. 
   The absorption-corrected 2--10~keV
  luminosity to infrared luminosity ratios of $F$05189--2524 during its ``high'' state is consistent with values observed in PG quasars.

\item The XIS spectrum of Mrk~273 contains a strong Fe~K$\alpha$ line
  and shows a change in the observed 2--10~keV spectrum relative to
  previous {\it Chandra} and {\em XMM-Newton} observations.  Mrk~273
  is marginally (1.8-$\sigma$) detected at high energies, with the
  HXD/PIN spectrum $\sim$5.4\% above the background.  A change in the
  covering fraction of the absorber best explains the spectral
  variations in Mrk~273, although a drop in the intrinsic AGN
  luminosity cannot be formally ruled out.  A column density of
  $\sim$10$^{24}$~cm$^{-2}$ is derived from the {\em Suzaku} data.
  The changes in spectral shape and covering fraction on a time scale
  of a few years suggest that the absorbing matter is $\sim$1 pc
  from the central source.  The Mrk~273 spectrum is best modeled by a
  $\sim$94\% covering fraction model.  The absorption-corrected
  2--10~keV luminosity to infrared luminosity ratios of Mrk~273 is
  consistent with values observed in PG quasars.

\item $F$08572+3915 is undetected in the {\it Suzaku} XIS and HXD/PIN
  observations.  The low X-ray count rate derived from unpublished
  {\it Chandra} observations, combined with mid-infrared observations, suggests that this source is highly obscured.

\item PKS~1345+12 is detected by the XIS.  No
  apparent flux or spectral variability is detected in its 0.5--10~keV
  spectrum relative to previous {\it Chandra} observations.  The net 
  15--40~keV HXD/PIN spectrum is only 1.1$\sigma$ above the total background, not
  strong enough for a meaningful full-band (XIS+HXD/PIN) spectral
  fitting.  Unlike other ULIRGs, the 0.5--2~keV spectrum of this
  source does not contain an obvious thermal MEKAL component.  Combining this
  result and the fact that the absorption-corrected 2--10~keV
  to infrared luminosity ratio of PKS 1345+12 is in agreement with those
  of PG quasars, the data suggest that the X-ray luminosity of this object is dominated by an AGN.

\item Arp~220 is detected by the XIS, but not by the HXD/PIN.  Its
  0.5--10~keV spectrum, including the iron complex, appears unchanged
  since previous {\it Chandra} observations.  The X-ray continuum
  emission is in agreement with the possibility that a highly obscured
  AGN is present.  This interpretation of the data is consistent with previous CO
  observations.  The measurements of the iron emission line at 6.7~keV can be reconciled by the ionized reflection model.
  
\item In all three cases where an optically thin thermal component is contributing to
  the soft X-ray emission detected by XIS
  ($F$05189--2524, Mrk~273, Arp~220), the temperature, $kT$
  $\sim$0.3--0.8~keV, is consistent with previous
  observations of ULIRGs.

\end{enumerate}

\acknowledgements

We are grateful to the anonymous referee for providing very useful comments and suggestions that have greatly improved this paper.  We would like to acknowledge Dr.\ Coleman Miller for useful
discussions.  We are also grateful for the extensive help on data
reduction provided by the NASA/HEASARC {\it Suzaku} GOF team.  The
data reduction for this paper made use of a module in the Beowulf
cluster (``the Borg'') in the Department of Astronomy, University of
Maryland.  This research also utilized the NASA/IPAC Extragalactic
Database (NED), which is operated by the Jet Propulsion Laboratory,
Caltech, under contract with NASA.  We acknowledge support from the
NASA/{\it Suzaku} Guest Observer Program under grant NNX06AI39G.  This
research has made use of data obtained from the {\it Suzaku} satellite, a collaborative mission between the space agencies of Japan (JAXA) and the USA (NASA).

\clearpage

%% Tables
%% Tables

\begin{deluxetable}{cccccccc}
\tabletypesize{\scriptsize}
\setlength{\tabcolsep}{0.03in}
\tablecolumns{8} \tablewidth{0pc}
  \tablecaption{The Sample}
  \tablehead{\colhead{Object}&\colhead{$z$}&\colhead{log F$_{\rm IR}$}&\colhead{$f_{25~\mu {\rm
    m}}$/$f_{60~\mu {\rm m}}$}&\colhead{Spectral}&\colhead{N$_{\rm H, Galactic}$}&\colhead{Scale}&\colhead{D$_{\rm L}$}\\
\colhead{}&\colhead{}&\colhead{[ergs/s/cm$^2$]}&\colhead{}&\colhead{Type}&\colhead{[10$^{20}$
  cm$^{-2}$]}&\colhead{[kpc/$\arcsec$]}&\colhead{[Mpc]}\\
\colhead{(1)}&\colhead{(2)}&\colhead{(3)}&\colhead{(4)}&\colhead{(5)}&\colhead{(6)}&\colhead{(7)}&\colhead{(8)}}
\startdata
F05189--2524&0.042&--8.87&0.25&S2/S1&1.92&0.71&173.3\\
F08572+3915&0.058&--9.18&0.23&LINER/S2&2.60&1.05&242.0\\
Mrk 273&0.038&--8.78&0.10&S2&1.10&0.70&156.3\\
PKS 1345+12&0.122&--9.65&0.35&S2/S1&1.90&2.05&531.8\\
Arp 220&0.018&--8.11&0.08&LINER&4.27&0.34&72.9\\
\enddata
\label{tab:sample}
\tablecomments{Col. (1): Object name. Col (2): redshift. Col. (3):
  logarithm of infrared (8--1000 $\mu$m) flux. Col. (4): the 25-to-60 $\mu$m {\it IRAS} flux ratio, a
  measure of the dust temperature.  Col. (5): galaxy spectral type.
  Col. (6): Galactic hydrogen column density \citep{nh}.  Col. (7): Physical
  size corresponding to 1$\arcsec$.  Col. (8): Luminosity distance.}
\end{deluxetable}

%% Tables

\begin{deluxetable}{cccccc}
\tabletypesize{\scriptsize}
\setlength{\tabcolsep}{0.03in}
\tablecolumns{6} \tablewidth{0pc}
  \tablecaption{{\it Suzaku} Observations}
  \tablehead{\colhead{Object}&\colhead{Observation}&\colhead{PI}&\colhead{Date}&\colhead{XIS Net
      Exposure}&\colhead{HXD Net Exposure}\\
\colhead{}&\colhead{ID}&\colhead{Name}&\colhead{[UT]}&\colhead{[ksec]}&\colhead{[ksec]}\\
\colhead{(1)}&\colhead{(2)}&\colhead{(3)}&\colhead{(4)}&\colhead{(5)}&\colhead{(6)}}
\startdata
F05189--2524&701097010&Veilleux; Anabuki&2006 April 10&78.2& 48.0\\
F08572+3915&701053010&Gallagher&2006 April 14&77.2 & 58.1\\
Mrk 273&701050010&Veilleux&2006 July 7&79.9 &  76.3\\
PKS 1345+12&702053010&Veilleux&2008 January 7&53.0&41.4\\
Arp 220&700006010&{\it Suzaku} SWG&2006 January 7&98.6 & 86.9\\
\enddata
\label{tab:obs}
\tablecomments{Col. (1): Object name. Col (2)--(3): {\it Suzaku}
  proposal number and principal investigator of archived data. 
  Col. (4): beginning observing date in UT.  Col. (5)--(6): net exposure
  time in kiloseconds after screening and HXD deadtime corrections.}

\end{deluxetable}

%% Tables

\begin{deluxetable}{ccccccccccccccccc}
\tabletypesize{\tiny}
\setlength{\tabcolsep}{0.02in}
\rotate 
\tablecolumns{17} \tablewidth{0pc}
  \tablecaption{Best-fit Parameters to {\it Suzaku} Spectra$^\dag$}
  \tablehead{\colhead{Object}&\colhead{Best-fit}&\colhead{kT}&\colhead{$\Gamma$}&\colhead{N$_{\rm H,~source}$}&\colhead{$f$}&\colhead{E$_{\rm line}$}&\colhead{$\sigma$}&\colhead{EW}&\colhead{$K$}&\colhead{$\chi ^2$/d.o.f.}&\colhead{Observed}&\colhead{Unabsorbed}&\colhead{Observed}&\colhead{Unabsorbed}&\colhead{Observed}&\colhead{Unabsorbed}\\
    \colhead{}&\colhead{Model}&\colhead{[keV]}&\colhead{}&\colhead{[10$^{23}$
    cm$^{-2}$]}&\colhead{}&\colhead{[keV]}&\colhead{[keV]}&\colhead{[keV]}&\colhead{}&\colhead{}&\colhead{F$_{\rm 0.5-2~keV}$}&\colhead{F$_{\rm 0.5-2~keV}$}&\colhead{F$_{\rm 2-10~keV}$}&\colhead{F$_{\rm 2-10~keV}$}&\colhead{F$_{\rm 12-40~keV}$}&\colhead{F$_{\rm 12-40~keV}$}\\
\colhead{(1)}&\colhead{(2)}&\colhead{(3)}&\colhead{(4)}&\colhead{(5)}&\colhead{(6)}&\colhead{(7)}&\colhead{(8)}&\colhead{(9)}&\colhead{(10)}&\colhead{(11)}&\colhead{(12)}&\colhead{(13)}&\colhead{(14)}&\colhead{(15)}&\colhead{(16)}&\colhead{(17)}}
\startdata
\cutinhead{XIS Only}\\
F05189--2524&A&0.71$^{+0.13}_{-0.08}$&2.68$^{+0.30}_{-0.13}$& --- &82.73&6.40$^{+0.14}_{-0.87}$&0.00$^{+0.82}_{-0.00}$&0.47$^{+0.75}_{-0.41}$&$2.20^{+0.48}_{-0.74} \times 10^{-5}$&185/123&0.66$^{+0.10}_{-0.10}$&0.70$^{+0.14}_{-0.10}$&1.03$^{+0.39}_{-0.38}$&1.03$^{+0.40}_{-0.40}$&---&---\\
%%F05189--2524&B&0.67$^{+0.07}_{-0.06}$&1.49$^{+0.15}_{-0.14}$& --- &---&6.40$^{+0.17}_{-0.14}$&0.25$^{+0.19}_{-0.08}$&1.84$^{+1.56}_{-1.45}$&194/124&0.63$^{+0.10}_{-0.09}$&0.63$^{+0.10}_{-0.11}$&1.01$^{+0.20}_{-0.25}$&1.01$^{+0.17}_{-0.24}$&---&---\\
Mrk~273&B&0.64$^{+0.04}_{-0.04}$&1.56$^{+0.15}_{-0.09}$&8.64$^{+1.71}_{-1.20}$&0.09&6.39$^{+0.03}_{-0.03}$&0.00$^{+0.10}_{-0.00}$&0.56$^{+0.15}_{-0.11}$&$3.65^{+1.84}_{-0.90} \times 10^{-4}$&233/187&1.17$^{+0.10}_{-0.13}$&8.98$^{+4.10}_{-2.05}$&4.29$^{+0.13}_{-1.67}$&20.44$^{+9.48}_{-4.84}$&---&---\\
PKS 1345+12&B&---&1.64$^{+0.09}_{-0.11}$&0.43$^{+0.03}_{-0.07}$&0.02&---&---&---&$3.13^{+0.52}_{-0.64} \times 10^{-4}$&164/156&0.37$^{+0.09}_{-0.09}$&5.93$^{+2.47}_{-3.73}$&9.53$^{+3.17}_{-4.06}$&11.91$^{+3.12}_{-4.64}$&---&---\\
Arp~220&C&0.67$^{+0.07}_{-0.04}$&1.93$^{+0.04}_{-0.07}$&0.07$^{+0.02}_{-0.02}$&---&6.70$^{+0.13}_{-0.08}$&0.05$^{+0.27}_{-0.05}$&0.42$^{+0.54}_{-0.32}$&$1.80^{+0.78}_{-0.40} \times 10^{-9}$ $^\ddag$&181/176&0.66$^{+0.10}_{-0.11}$&1.85$^{+0.75}_{-0.40}$&0.98$^{+0.24}_{-0.42}$&1.04$^{+0.72}_{-0.35}$&---&---\\
\cutinhead{Combined XIS+HXD/PIN Spectra}\\
Mrk~273&B&0.64$^{+0.04}_{-0.04}$&1.40$^{+0.10}_{-0.11}$&14.09$^{+1.63}_{-1.53}$&0.05&6.38$^{+0.03}_{-0.04}$&0.03$^{+0.08}_{-0.03}$&0.68$^{+0.18}_{-0.15}$&$6.14^{+2.39}_{-1.77} \times 10^{-4}$&268/199&1.15$^{+0.12}_{-0.10}$&14.50$^{+5.32}_{-3.98}$&4.68$^{+0.75}_{-1.53}$&43.19$^{+16.03}_{12.09}$&66.59$^{+22.43}_{-35.53}$&76.27$^{+28.79}_{-21.45}$\\
\hline
\enddata

\tablecomments{%Model: Absorption$_{\rm Galactic} \times [{\rm
    %MEKAL} + {\rm Partial Covering Absorption_{source}} \times ({\rm PL +
    %Line})]$ + Contributions from nearby sources, where MEKAL is the Mewe, Kaastra, \& Liedahl thermal
  %plasma model (see the XSPEC manual for details), PL is a power-law
  %model representing the AGN, Line is an emission line with a
  %Gaussian profile.  
  $^\dag$ F08572+3915 is undetected by the XIS and is thus not included in this table.  $^\ddag$ Since the spectrum is pure reflection, the normalization factor is of the reflection component.  Col. (1): Object name.  Col. (2):  favored model -- (A) reflection plus power law model, (B) scattering model, (C) ionized reflection model.  Col (3): temperature of the
  thermal starburst component from the MEKAL model. Col. (4):
  slope to the power-law model. Col. (5): absorption within the
  source.  Col. (6): the ratio of the normalization parameter between the reflected (or scattered) component to the direct component.  Col. (7): rest energy of the emission line.
  Col. (8): line width of the emission line.  Col. (9): equivalent
  width of the emission line.  Col. (10): normalization factor of the direct component in photons cm$^{-2}$ s$^{-1}$ at 1~keV.  Col. (11): $\chi ^2$ fitting statistics
  per degree of freedom.  
  Col. (12)--(17): flux in units of 10$^{-13}$ ergs s$^{-1}$ cm $^{-2}$.}

\label{tab:fitxis}
\label{tab:fithxd}
\end{deluxetable}

%% Tables

\begin{deluxetable}{cccccccccccccccccc}
\tabletypesize{\tiny}
\setlength{\tabcolsep}{0.01in}
\rotate 
\tablecolumns{18} \tablewidth{0pc}
  \tablecaption{Best-fit Parameters to Spectra of F05189--2524 and Mrk
    273 from Multiple Epochs}
  \tablehead{\colhead{Telescope}&\colhead{kT}&\colhead{$\Gamma$}&\colhead{$f$}&\colhead{N$_{\rm H,~source~1}$}&\colhead{$f_{\rm cover~1}$}&\colhead{N$_{\rm H,~source~2}$}&\colhead{$f_{\rm cover~2}$}&\colhead{E$_{\rm line}$}&\colhead{$\sigma$}&\colhead{EW}&\colhead{$K$}&\colhead{Observed}&\colhead{Unabsorbed}&\colhead{Observed}&\colhead{Unabsorbed}&\colhead{Observed}&\colhead{Unabsorbed}\\
    \colhead{}&\colhead{[keV]}&\colhead{}&\colhead{}&\colhead{[10$^{23}$
    cm$^{-2}$]}&\colhead{}&\colhead{[10$^{23}$
    cm$^{-2}$]}&&\colhead{[keV]}&\colhead{[keV]}&\colhead{[keV]}&\colhead{}&\colhead{F$_{\rm 0.5-2~keV}$}&\colhead{F$_{\rm 0.5-2~keV}$}&\colhead{F$_{\rm 2-10~keV}$}&\colhead{F$_{\rm 2-10~keV}$}&\colhead{F$_{\rm 15-40~keV}$}&\colhead{F$_{\rm 15-40~keV}$}\\
\colhead{(1)}&\colhead{(2)}&\colhead{(3)}&\colhead{(4)}&\colhead{(5)}&\colhead{(6)}&\colhead{(7)}&\colhead{(8)}&\colhead{(9)}&\colhead{(10)}&\colhead{(11)}&\colhead{(12)}&\colhead{(13)}&\colhead{(14)}&\colhead{(15)}&\colhead{(16)}&\colhead{(17)}&\colhead{(18)}}
\startdata
\cutinhead{F05189--2524: scattering model, change in column density ($\chi ^2$/d.o.f. = 640/503)}\\
XMM
(2001)&0.23$^{+0.02}_{-0.03}$&1.91$^{+0.08}_{-0.05}$&0.02&0.58$^{+0.03}_{-0.03}$&---&---&---&6.37$^{+0.27}_{-0.25}$&0.73$^{+0.31}_{-0.16}$&0.64$^{+0.98}_{-0.63}$&$1.42^{+0.17}_{-0.12} \times 10^{-3}$&1.02$^{+0.10}_{-0.11}$&30.00$^{+1.29}_{-17.53}$&29.38$^{+5.42}_{-4.35}$&42.03$^{+6.58}_{-1.79}$&---&---\\
Chandra
(2001)&0.23&1.91&0.02&0.90$^{+0.05}_{-0.05}$&---&---&---&6.37&0.73&0.64&$1.42 \times 10^{-3}$&0.77$^{+0.08}_{-0.08}$&30.00&25.67$^{+3.33}_{-5.79}$&42.03&---&---\\
Chandra
(2002)&0.23&1.91&0.02&0.71$^{+0.04}_{-0.03}$&---&---&---&6.37&0.73&0.64&$1.42 \times 10^{-3}$&0.87$^{+0.09}_{-0.09}$&30.00&27.69$^{+3.66}_{-7.16}$&42.03&---&---\\
Suzaku
(2006)&0.23&1.91&0.02&21.55$^{+2.18}_{-1.71}$&---&---&---&6.37&0.73&0.64&$1.42 \times 10^{-3}$&0.70$^{+0.10}_{-0.08}$&30.00&1.45$^{+0.28}_{-0.25}$&42.03&---&---\\
\cutinhead{F05189--2524: reflection dominated, change in AGN luminosity ($\chi ^2$/d.o.f. = 659/502)}\\
XMM
(2001)&0.24$^{+0.02}_{-0.01}$&1.79$^{+0.07}_{-0.13}$&0.02&0.59$^{+0.05}_{-0.05}$&---&---&---&---&---&---&$1.27^{+0.24}_{-0.21} \times 10^{-3}$&1.02$^{+0.12}_{-0.10}$&27.02$^{+1.55}_{-18.21}$&29.92$^{+3.15}_{-4.03}$&42.80$^{+6.03}_{-0.50}$&---&---\\
Chandra
(2001)&0.24&1.79&0.02&0.93$^{+0.09}_{-0.09}$&---&---&---&---&---&---&$1.27 \times 10^{-3}$&0.77$^{+0.10}_{-0.11}$&27.02&25.90$^{+2.72}_{-3.03}$&42.80&---&---\\
Chandra
(2002)&0.24&1.79&0.02&0.72$^{+0.07}_{-0.06}$&---&---&---&---&---&---&$1.27 \times 10^{-3}$&0.87$^{+0.10}_{-0.10}$&27.02&28.16$^{+3.14}_{-3.82}$&42.80&---&---\\
Suzaku
(2006)&0.24&1.79&6.82&---&---&---&---&6.32$^{+0.17}_{-0.22}$&0.00$^{+0.46}_{-0.00}$&1.95$^{+3.64}_{-1.00}$&$2.18^{+0.30}_{-0.36} \times 10^{-5}$&0.69$^{+0.09}_{-0.10}$&0.75$^{+0.10}_{-0.10}$&1.06$^{+0.33}_{-0.14}$&1.06$^{+0.36}_{-0.11}$&---&---\\
\cutinhead{Mrk~273: change in covering fraction ($\chi ^2$/d.o.f. = 521/376)}\\
Chandra (2000)&0.79$^{+0.03}_{-0.03}$&1.73$^{+0.04}_{-0.02}$&---&15.91$^{+0.86}_{-0.73}$&0.80$^{+0.01}_{-0.02}$&3.00$^{+0.23}_{-0.20}$&0.96$^{+0.01}_{-0.01}$&6.40$^{+0.01}_{-0.13}$&0.00$^{+0.06}_{-0.00}$&0.25$^{+0.04}_{-0.07}$&$2.24^{+0.20}_{-0.11} \times 10^{-3}$&0.73$^{+2.92}_{-0.15}$&47.29$^{+4.25}_{-2.26}$&7.49$^{+3.06}_{-2.36}$&84.10$^{+7.89}_{-4.48}$&---&---\\
XMM(2002)&0.79&1.73&---&15.91&0.94$^{+0.01}_{-0.01}$&3.00&0.81$^{+0.02}_{-0.02}$&6.40&0.00&0.25&$2.24 \times 10^{-3}$&0.85$^{+0.05}_{-0.27}$&47.29&4.09$^{+0.55}_{-1.87}$&84.10&---&---\\
Suzaku(2006)&0.79&1.73&---&15.91&0.94$^{+0.01}_{-0.01}$&3.00&0.69$^{+0.02}_{-0.02}$&6.40&0.00&0.25&$2.24 \times 10^{-3}$&1.11$^{+0.03}_{-0.42}$&47.29&4.36$^{+0.30}_{-1.96}$&84.10&67.35$^{+22.51}_{-40.50}$&78.48$^{+8.40}_{-4.45}$\\
\hline
\enddata
\tablecomments{%%Model: Absorption$_{\rm Galactic} \times [{\rm
%%    MEKAL} + {\rm Partial Covering Absorption_{source}} \times ({\rm PL +
%%    Line})]$ + Contributions from nearby sources, where MEKAL is the Mewe, Kaastra, \& Liedahl thermal
%%  plasma model (see the XSPEC manual for details), PL is a power-law
%%  model representing the AGN, Line is an emission line with a
%%  Gaussian profile.  
Col. (1): Spacecraft used and year of the observation. Col (2): temperature of the
  thermal starburst component from the MEKAL model. Col. (3):
  slope to the power-law model.  Col. (4): the ratio of the normalization parameters between the reflected (or scattered) component to the direct component.  Col. (5)--(8): absorption within the
  source and corresponding covering fraction for the covering fraction
  absorber model.  Col. (9): rest energy of the emission line.
  Col. (10): line width of the emission line.  Col. (11): equivalent
  width of the emission line.  Col. (12): normalization factor of the direct component in photons cm$^{-2}$ s$^{-1}$.  Col. (13)--(18): flux in units of
  10$^{-13}$ ergs s$^{-1}$ cm$^{-2}$.}

\label{tab:fitmulti}
\end{deluxetable}

%% Figures
\clearpage

\end{document}